\documentclass[graybox]{svmult}

\usepackage{amsmath,bm,cite}
\usepackage{helvet}         % selects Helvetica as sans-serif font
\usepackage{courier}        % selects Courier as typewriter font
\usepackage{type1cm}        % activate if the above 3 fonts are
\usepackage{graphicx}        % standard LaTeX graphics tool
\usepackage{multicol,multirow}        % used for the two-column index
\usepackage[bottom]{footmisc}% places footnotes at page bottom
\usepackage[usenames,dvipsnames,table]{xcolor}
\usepackage{colortbl}
\usepackage{pbox}
\usepackage{chemarr,mathtools}

\usepackage[asymmetric,left=2.2cm,textwidth=17cm,textheight=24cm,top=2cm]{geometry}

\usepackage{titlesec}
\titleformat*{\section}{\Large\bfseries\sffamily}
\titleformat*{\subsection}{\large\bfseries\sffamily}
\titleformat*{\subsubsection}{\normalfont\bfseries\sffamily}

\newcommand{\avg}[1]{\left\langle #1 \right\rangle}
\newcommand{\ovl}[1]{\overline{#1}}

\newcommand{\inn}{\textrm{on}}
\newcommand{\out}{\textrm{off}}

\newcommand{\btil}{b^o}
\newcommand{\betil}{\beta_o}

\newcommand{\maxx}{\textrm{max}}
\newcommand{\minn}{\textrm{min}}

\newcommand{\til}[1]{\widetilde{ #1 }}

\newcommand{\uudl}[1]{\underline{\underline{#1}}}

\renewcommand{\thesection}{\Roman{section}}

\begin{document}

\title*{\textsf{Kinetic modelling of competition and depletion of shared miRNAs by competing endogenous RNAs}}
\titlerunning{Modelling competing endogenous RNAs}
% your contribution title if the original one is too long
\author{Araks Martirosyan, Marco Del Giudice, Chiara Enrico Bena, Andrea Pagnani, Carla Bosia, Andrea De Martino}
\authorrunning{Martirosyan {\it et al.}}
\institute{Araks Martirosyan \at Laboratory of Glia Biology, VIB-KU Leuven Center for Brain and Disease Research and KU Leuven Department of Neuroscience, 
O\&N4 Herestraat 49 box 602, 3000 Leuven (Belgium)\\ \email{araks.martirosyan@kuleuven.vib.be}
\and Marco Del Giudice, Chiara Enrico Bena, Andrea Pagnani and Carla Bosia\at DISAT, Politecnico di Torino, corso Duca degli Abruzzi 24, 10129 Turin (Italy) and Italian Institute for Genomic Medicine, via Nizza 52, 10126 Turin (Italy) \\ \email{marco.delgiudice@polito.it; chiara.enrico@polito.it; andrea.pagnani@polito.it; carla.bosia@polito.it}\
\and Andrea De Martino\at Soft \& Living Matter Lab, CNR-NANOTEC, Rome (Italy) and Italian Institute for Genomic Medicine, via Nizza 52, 10126 Turin (Italy) \\
\email{andrea.demartino@roma1.infn.it}
}
%
% Use the package "url.sty" to avoid
% problems with special characters
% used in your e-mail or web address
%
\maketitle

\abstract{Non-conding RNAs play a key role in the post-transcriptional regulation of mRNA translation and turnover in eukaryotes. miRNAs, in particular, interact with their target RNAs through protein-mediated, sequence-specific binding, giving rise to extended and highly heterogeneous miRNA-RNA interaction networks. Within such networks, competition to bind miRNAs can generate an effective positive coupling between their targets. Competing endogenous RNAs (ceRNAs) can in turn regulate each other through miRNA-mediated crosstalk. Albeit potentially weak, ceRNA interactions can occur both dynamically, affecting e.g. the regulatory clock, and at stationarity, in which case ceRNA networks as a whole can be implicated in the composition of the cell's proteome. Many features of ceRNA interactions, including the conditions under which they become significant, can be unraveled by mathematical and {\it in silico} models. We review the understanding of the ceRNA effect obtained within such frameworks, focusing on the methods employed to quantify it, its role in the processing of gene expression noise, and how network topology can determine its reach.}

%\keywords{miRNA, ceRNA, competition, sponging, mathematical modeling}

\section{Introduction}
\label{sec:1}

microRNAs (miRNAs) --short, endogenous, noncoding RNAs that operate post-transcriptionally via sequence-specific binding to target RNAs-- are increasingly recognized as key actors in the regulation of eukaryotic gene expression \cite{bartel,flynt,cech,gurt,metaz}. Following transcription (from either introns of protein-coding genes or from miRNA-specific genes) and maturation, miRNAs get incorporated into specialized, multiprotein complexes known as RISCs (short for RNA-induced silencing complexes) \cite{risc}. Once within a RISC, the miRNA provides the pattern to bind specific sites called miRNA response elements (MREs) found on their target RNAs \cite{chan,why}. Effective base pairing typically requires 6- to 9-nucleotide complementarity, and leads to negative gene expression control through either mRNA destabilization or translational repression  \cite{chek,jona,djur}. The fact that miRNA expression is significantly tissue-specific places miRNAs at the center of the regulatory layer that controls the composition of the protein repertoire and cell type specificity \cite{bart,liang,fran,eber}. Still, many aspects of miRNA biology  suggest that this role might be exerted through a broader and  more complex, yet possibly more subtle, class of mechanisms. 

In first place, miRNAs appear to be highly conserved in vertebrates and invertebrates, and their mRNA target structure displays a significant degree of conservation in higher organisms \cite{bere,josh}. For instance, more than half of human genes are conserved miRNA targets, including a large number of weak-interacting sites that appear to be under selective pressure to be maintained \cite{frie}. Such a strong degree of conservation suggests that protein levels may need to be fine tuned within extremely precise ranges \cite{baek}. Quantitative studies together with the statistical overrepresentation of noise-buffering motifs within the miRNA-RNA network indeed supports this idea \cite{shim,tsang,reda}, and recent experiments have confirmed miRNA's ability to stabilize output levels for lowly expressed proteins \cite{sici}. Yet, the amount of noise reduction that can be achieved even in optimal conditions does not seem to justify a view of noise suppression as the key evolutionary driver for a significantly conserved miRNA targeting pattern \cite{wangs,das,ober,schm}. 

Secondly, miRNA targets are known to include, together with messenger RNAs, a host of ncRNA species like lncRNAs as well as pseudogenes \cite{guil,hans,eber2}. On one hand, miRNA sponging by ncRNAs can clearly be critical in determining both miRNA levels and their potential for translational repression. On the other, it substantially increases the complexity of the network of miRNA-RNA interactions. It is now clear that each long RNA molecule can typically be targeted by multiple miRNAs, while every miRNA can interact with a very large number of distinct RNAs, generating an  extended interaction network stretching across the entire transcriptome \cite{suma,helw,kimd,zavo}. Now the ability of miRNAs to regulate gene expression is ultimately linked to the overall target availability, and tends to get weaker as the number of targets (more precisely, of potential binding sites) increases, the so-called `dilution' effect \cite{arve}. This leaves room to search for alternative mechanisms through which miRNAs could exert a regulatory function, even at the non-local (up to system-scale) level.

The heterogeneity of the miRNA-RNA network and the fact that repression potential depends tightly on molecular levels suggest that competition to bind miRNAs might be a contributing factor in the establishment of robust protein profiles \cite{levine,fzor}. In rough terms, the essence of the so-called ceRNA hypothesis (whereby `ceRNA' stands for `competing endogenous RNA') is that, due to a cross-correlation of molecular levels, competition can induce an effective positive coupling between miRNA targets, such that a perturbation affecting the level of one target could be broadcast to its competitor via the subsequent shift in miRNA availability \cite{salm}. In this respect, one might say that RNAs form a sort of `molecular ecosystem', where mutual dependencies can be established post-transcriptionally via miRNA-mediated interactions driven by competition. The ceRNA scenario has received much attention since its formulation, both {\it ex vivo} and in synthetic systems (see e.g.   \cite{tay,vano,karr,tayy,yuany,sgro}). Effective interactions coupling RNAs targeted by the same miRNAs (which can be probed e.g. by over-expressing miRNAs or targets) are now known to be implicated in a variety of processes, from development and differentiation \cite{fati}, to stress response \cite{stress} and disease \cite{alva,anas}, and have been investigated in connection to their perspective therapeutic usefulness \cite{sanc}. 

Still, it has also become clear that the theoretical appeal of the ceRNA effect is not easily translated into quantitative understanding. A key issue is that of fine tuning. Several conditions clearly factor in the emergence of the ceRNA scenario. The possibility to turn competition between miRNA targets into an effective positive coupling between them presupposes for instance a cross-coordination of molecular levels, as a large excess (resp. scarcity) of miRNAs with respect to targets or binding sites will necessarily result into a completely repressed (resp. unrepressed) profile \cite{jens,denzler}. The ceRNA scenario would naturally become less realistic if kinetic parameters had to be tightly tuned in order to allow for ceRNA crosstalk conditions to arise. In addition, experiments suggest that a relatively small number of targets are usually sensitive to modulation in miRNA availability. Moreover, which targets are responsive depends on miRNA levels \cite{alau,boss,denz}. The emergent selectivity and adaptability of ceRNA interactions should be reconciled with the heterogeneity observed in the miRNA-RNA interaction network in which each miRNA can regulate up to hundreds of targets. 

Mathematical and {\it in silico} models developed in recent years have shed light on several of these issues and revealed many unexpected traits \cite{wang,laix}. This chapter aims at reviewing the methods employed and the key features of the ceRNA scenario that  such studies suggest. 

Our starting point is a generic, minimal deterministic mathematical model of post-transcriptional regulation whose steady states can be fully characterized analytically and numerically. Despite its roughness, it allows to precisely quantify the sensitivity of a ceRNA to alterations in the level of one of its competitors, sufficing to capture many of the central characteristics of miRNA-based regulation from basic assumptions about the underlying processes. In particular, miRNA-ceRNA interaction strengths and silencing/sequestration mechanisms emerge, together with the relative abundance of regulators and targets, as key factors for the onset and character of ceRNA crosstalk, including its selectivity. Moreover, heterogeneities in kinetic parameters as well as in miRNA-ceRNA interaction topology are major drivers of ceRNA crosstalk in a broad range of parameter values. The picture obtained at stationarity can be extended to out-of-equilibrium regimes. In particular, one can characterize a `dynamical' ceRNA effect, which can be stronger than the equilibrium one, as well as the typical timescales required to reach stationary crosstalk. 

Passing from a deterministic to a stochastic description, one can address the behaviour of fluctuations in molecular levels and evaluate the ability of miRNA-based regulatory elements to process noise. We will show in particular that the ceRNA mechanism can provide a generic pathway to the reduction of intrinsic noise both for individual proteins and for complexes formed by sub-units sharing a miRNA regulator (which might explain why interacting proteins are frequently regulated by miRNA clusters). The processing of extrinsic (transcriptional) noise is more involved. While ceRNA crosstalk is generically hampered by it, specific patterns of transcriptional correlations can actually result in enhanced noise buffering and in the emergence of complex (e.g. bistable) expression patterns. On the other hand, one can quantify the physical limits to crosstalk intensity by considering how different sources of noise affect it. It turns out that the size of target derepression upon the activation of its competitor is a crucial determinant of the  strength of miRNA-mediated ceRNA regulation. When it is sufficiently large, post-transcriptional crosstalk can be as effective as direct transcriptional regulation in controlling expression levels. In specific cases, ceRNA crosstalk may even represent the most effective mechanism to tune gene expression. 

An especially important question (and a difficult one, in view of the fact that the effect can be rather modest) concerns the quantification of ceRNA crosstalk intensity, and specifically the identification of unambiguous crosstalk markers that can be validated both experimentally and through the analysis of transcriptional data. We shall examine a few alternatives that have been employed, highlighting the different motivations underlying their use, their physical meaning and their respective limitations.

\section{Models and methods}
\label{sec:2}

\subsection{Deterministic model}

The simplest mathematical representation of the dynamics of $N$ ceRNA species and $M$ miRNA species interacting in a miRNA-ceRNA network is based on deterministic mass-action kinetics. We shall denote by $m_i$ the level of ceRNA species $i$ (with $i$ ranging from 1 to $N$), by $\mu_a$ the level of  miRNA species $a$ (ranging from 1 to $M$), and by $c_{ia}$ the levels of miRNA-ceRNA complexes. Based on experimental evidence, one can assume that all miRNA molecules are `active', i.e. bound to an Argonaute protein and ready to attach to a target ceRNA. This allows to discard the kinetic steps leading to the formation of the RNA-induced silencing complex (RISC). In such conditions, concentration variables evolve in time due to 
\begin{enumerate}
\item synthesis and degradation events,
\item complex binding and unbinding events,
\item the processing of complexes.
\end{enumerate}
The latter in turn can follow two distinct pathways: a catalytic one, leading to the degradation of the ceRNA with the re-cycling of the miRNA; and a stoichiometric one, where both molecules are degraded, possibly after sequestration into P-bodies \cite{vale,bacc}. The relevant processes (see Fig. 1A and B for a sketch) are therefore 
\begin{equation}\label{processes}
\arraycolsep=1pt\def\arraystretch{1.8}
\begin{array}{r@{}l}
&{}\emptyset \xrightleftharpoons[d_i]{b_i} m_i ~~~~~~~~~~~~
\emptyset \xrightleftharpoons[\delta_a]{\beta_a} \mu_a ~~~~~~~~~~~~
\mu_a+m_i \xrightleftharpoons[k_{ia}^-]{k_{ia}^+} c_{ia}\\
&{}c_{ia} \xrightharpoonup{\sigma_{ia}} \emptyset ~~~~~~~~~~~~
c_{ia} \xrightharpoonup{\kappa_{ia}} \mu_a 
\end{array}~~~~~~~.
\end{equation}
Correspondingly, the mass action kinetic equations take the form (see e.g. \cite{figl,bosi,meht})
\begin{equation}\label{uno}
\arraycolsep=1pt\def\arraystretch{2}
\begin{array}{r@{}l}\frac{d m_i}{dt}&{}=b_i-d_i m_i-\sum_a k_{ia}^+ m_i\mu_a +\sum_a k_{ia}^- c_{ia}~~, \\
\frac{d \mu_a}{dt}&{}=\beta_a-\delta_a \mu_a-\sum_i k_{ia}^+ m_i\mu_a + \sum_{i}(k_{ia}^-+\kappa_{ia})c_{ia}~~,\\
\frac{d c_{ia}}{dt}&{}= k_{ia}^+ m_i \mu_a-(\sigma_{ia}+\kappa_{ia}+k_{ia}^-)c_{ia}~~,
\end{array}
\end{equation}
where the physical meaning of parameters is summarized in Table \ref{tab1} and where the indices $i$ and $a$ range from $1$ to $N$ and from $1$ to $M$, respectively. 
\begin{table}[b]
\caption{Variables and parameters appearing in the basic model, Eq. (\ref{uno}). Note that the levels of molecular species can be specified by copy numbers (as indicated below) as well as by (continuous) concentrations, depending on whether the modeling framework is stochastic (see Section \ref{stochastic}) or  deterministic (as in Eq. (\ref{uno})), respectively. \label{tab1}}
\begin{tabular}{p{1.5cm}p{2.4cm}p{7.5cm}}
\hline\noalign{\smallskip}
Variable & Units & Description   \\
\noalign{\smallskip}\svhline\noalign{\smallskip}
$m_i$ & molecules & Number of free copies of ceRNA species $i$\\
$\mu_a$ & molecules & Number of free copies of miRNA species $a$\\
$c_{ia}$ & molecules & Number of copies of $i-a$ complex\\ [0.2cm]
\hline\noalign{\smallskip}
Parameter & Units & Description   \\
\noalign{\smallskip}\svhline\noalign{\smallskip}
$b_i$ & molecule min$^{-1}$ & Transcription rate of ceRNA species $i$\\
$d_i$ & min$^{-1}$ & Degradation rate of ceRNA species $i$\\
$\beta_a$ & molecule min$^{-1}$ & Transcription rate of miRNA species $a$\\
$\delta_a$ & min$^{-1}$ & Degradation rate of miRNA species $i$\\
$k_{ia}^+$ & molecule$^{-1}$ min$^{-1}$ & $i-a$ complex association rate\\
$k_{ia}^-$ & min$^{-1}$ & $i-a$ complex dissociation rate\\
$\kappa_{ia}$ & min$^{-1}$ & Catalytic decay rate (with miRNA re-cycling) of $i-a$ complex\\
$\sigma_{ia}$ & min$^{-1}$ & Stoichiometric decay rate (without miRNA re-cycling) of $i-a$ complex\\
\noalign{\smallskip}\hline\noalign{\smallskip}
\end{tabular}
\end{table}
\begin{figure}[t]
\begin{center}
\includegraphics[width=16cm]{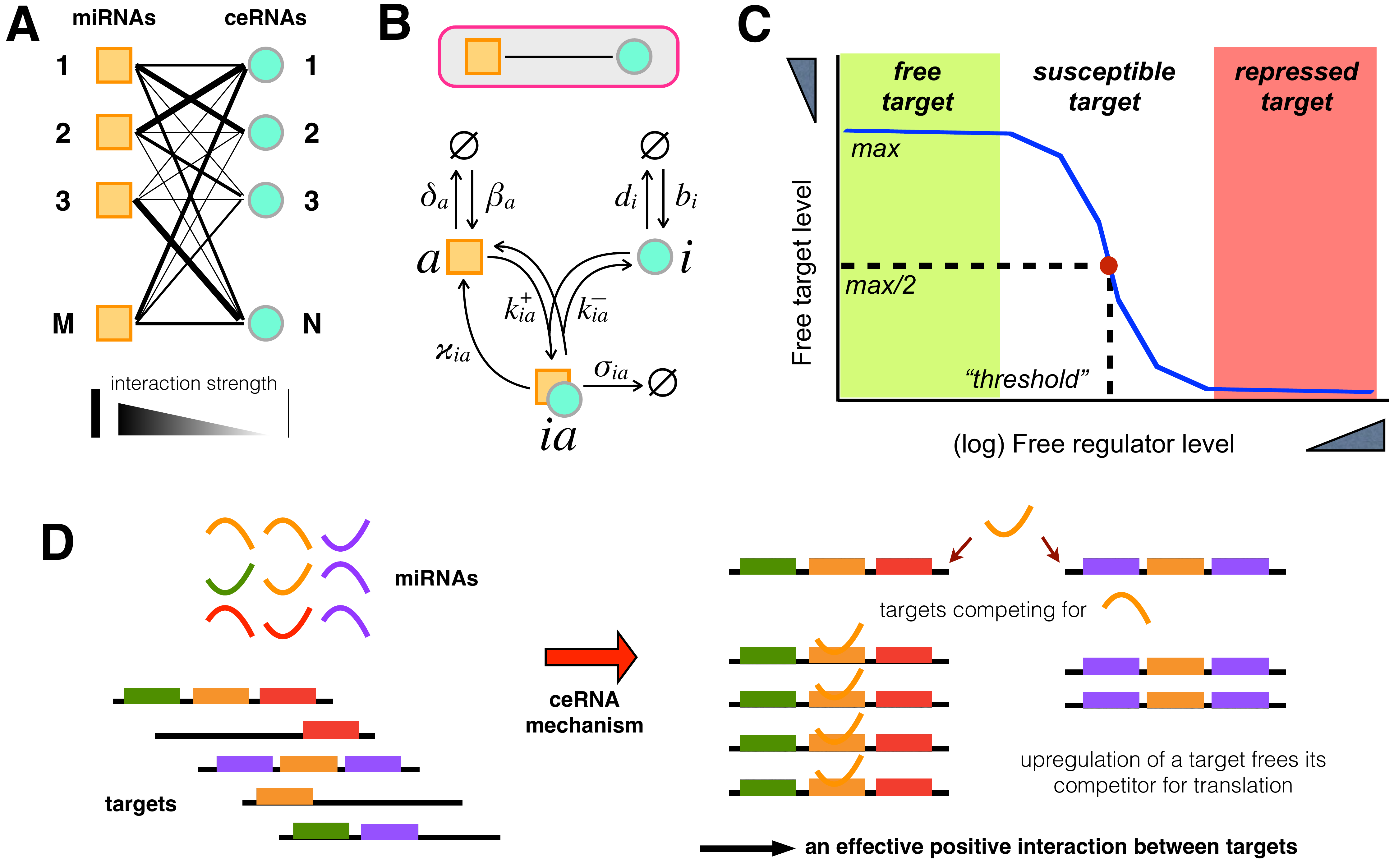}
\end{center}
\caption{{\bf (A)} Sketch of an interaction network formed by miRNAs and their targets (ceRNAs). The network is a weighted bipartite graph. Line thickness is proportional to the coupling strength (i.e. to the miRNA-ceRNA binding affinity). {\bf (B)} Sketch of the individual processes lumped in each interaction represented in (A). Details of reactions and rates are given in Eq. (\ref{processes}). {\bf (C)} Sketch of the behaviour of the level of free targets (ceRNA or miRNA) as a function of the level of free regulators  (miRNA or ceRNA, respectively). {\bf (D)} Sketch of the ceRNA mechanism: competition to bind a miRNA can induce an effective positive coupling between its targets.}
\label{Fig1}       
\end{figure}
For several purposes it is useful to introduce the ``stoichiometricity ratio''
\begin{equation}\label{sr}
\alpha_{ia}=\frac{\sigma_{ia}}{\sigma_{ia}+\kappa_{ia}}
\end{equation} 
quantifying the probability that the $i-a$ complex is processed without miRNA re-cycling.

\paragraph{{\bf \textsf{Note}} }

The model just described, that is the one on which we will mostly focus, is limited to miRNAs and ceRNAs and excludes, for instance, upstream regulators (e.g. transcription factors, TFs) and downstream products (e.g. proteins). Integrating some of these ingredients is however  straightforward and it has been done in the literature. For instance, upstream TFs independently regulating the synthesis of ceRNAs and miRNA can be accounted for by assuming that transcription requires the cooperative binding of $H$ TF molecules for each of the RNA species involved (labeled $\ell$, including both miRNAs and ceRNAs). Denoting by $k_{\inn}$ and $k_{\out}$ the binding and unbinding rates of TFs to DNA, respectively, the fractional occupancies of TF binding sites on the DNA evolve as
\begin{equation} 
\frac{d n_{\ell}}{dt} = k_{\inn}(1-n_{\ell}) f_\ell^H - k_{\out} n_{\ell}~~,
\end{equation}
where $n_{\ell}$ ($0\leq n_{\ell}\leq 1$) stands for the probability that the binding site for the TF controlling the transcription of species $\ell$ is occupied and $f_\ell$ stands for the level of the TF controlling species $\ell$. In most cases, the variables $n_\ell$ will equilibrate on timescales much shorter than those characterizing the dynamics of molecular levels \cite{alon}. In such conditions, each $n_{\ell}$ can be thought to take on its stationary value, i.e.
\begin{equation}
\avg{n_\ell}=\frac{f_\ell^H } {f_\ell^H + K^H}~~~~~,~~~~~K=\left(\frac{k_{\out}}{k_{\inn}}\right)^{1/H}~~.
\end{equation}
Such occupancies in turn modulate the transcription rates appearing in (\ref{uno}). In particular, the effective transcription rate of ceRNA (resp. miRNA) species $i$ (resp. $a$) becomes $b_{i,{\rm eff}}= b_i\,\avg{n_i}$ (resp. $\beta_{a,{\rm eff}}= \beta_a\,\avg{n_a}$) \cite{prob}.

An extension of (\ref{uno}) including downstream species (proteins) is briefly discussed in Sec. \ref{prots}.

\subsection{Analysis of the steady state: threshold behaviour and competition-induced responses}

At steady state, molecular populations evolving according to Eqs (\ref{uno}) are given by the solutions of
\begin{equation}\label{eq:steadystateM}
\arraycolsep=1pt\def\arraystretch{2}
\begin{array}{r@{}l}
\avg{m_i} &{}= \frac{b_i + \sum_a k_{ia}^- \avg{c_{ia}} }{d_i + \sum_a k_{ia}^+ \avg{\mu_a} }~~ \\ 
\avg{\mu_a} &{}= \frac{\beta_a  + \sum_i (k_{ia}^- + \kappa_{ia}) \avg{c_{ia}} }{ \delta_a + \sum_i k^+_{ia} \avg{m_i} }~~\\
\avg{c_{ia}} &{}= \frac{k^+_{ia} \avg{\mu_a} \, \avg{m_i} }{\sigma_{ia} + \kappa_{ia} + k^-_{ia}}~~
%\label{eq:steadystateMEND}
\end{array}~~.
\end{equation}
(We shall henceforth represent the steady state level of species $x$ by angular brackets, i.e. $\avg{x}$.) These conditions have been rigorously shown to describe the unique, asymptotically stable steady state of (\ref{uno}) \cite{flon}. Eqs (\ref{eq:steadystateM}) provide a full description of the molecular network in terms of the populations of all species at sufficiently long times, given all kinetic parameters, and are easily solved numerically for any $N$ and $M$. It is however possible to get a mathematical intuition about how miRNAs affect ceRNA levels at stationarity by eliminating complexes (i.e. $\avg{c_{ia}}$) from (\ref{eq:steadystateM}). This allows to re-cast the steady-state in terms of miRNA and ceRNA levels only. Specifically, one gets 
\begin{equation}\label{ss}
\arraycolsep=1pt\def\arraystretch{1.8}
\begin{array}{r@{}l}
\avg{m_i} &{}= \frac{m_i^\star}{1+\sum_a \mu_a/\mu_{0,ia}}~~, \\ 
\avg{\mu_a} &{}= \frac{\mu_a^\star}{1+\sum_i m_i/m_{0,ia}}~~,
\end{array}
\end{equation}
where $m_i^\star\equiv b_i/d_i$ and $\mu_a^\star=\beta_a/\delta_a$ stand for the maximum values achievable by ceRNA and miRNA levels at stationarity, while
\begin{equation}\label{thresholds}
\arraycolsep=1pt\def\arraystretch{1.8}
\begin{array}{r@{}l}
m_{0,ia} &{}= \frac{\delta_a}{k_{ia}^+}\left(1+\frac{k_{ia}^-+\kappa_{ia}}{\sigma_{ia}}\right)~~, \\ 
\mu_{0,ia} &{}= \frac{d_i}{k_{ia}^+}\left(1+\frac{k_{ia}^-}{\sigma_{ia}+\kappa_{ia}}\right)~~
\end{array}
\end{equation}
represent `reference' concentrations that depend on the specific miRNA-ceRNA pair. For sakes of simplicity, we shall refer to these values as ``thresholds''. The gist of (\ref{ss}) is the following (see Fig. \ref{Fig1}C) \cite{figl}:
\begin{description}
\item[{\bf Free or unrepressed regime}~:]If the levels of all miRNA species interacting with ceRNA $i$ are sufficiently low (specifically, much lower than the respective thresholds $\mu_{0,ia}$, so that $\sum_a \mu_a/\mu_{0,ia}\ll 1$), then the steady-state level of ceRNA $i$ will be very close to the maximum possible, $m_i^\star$. In such conditions, ceRNA species $i$ will be roughly insensitive to changes in miRNA levels. We call this the `unrepressed' or `free' regime for ceRNA $i$.  
\item[{\bf Susceptible regime}~:]As the quantity $\sum_a \mu_a/\mu_{0,ia}$ increases, e.g. following an increase in the level of one or more miRNA species, $\avg{m_i}$ deceases in a sigmoidal fashion. This occurs most notably when $\sum_a \mu_a/\mu_{0,ia}\simeq 1$ (corresponding, for $M=1$, to a miRNA level close to the threshold value $\mu_{0,ia}$). Here ceRNA $i$ is very sensitive to a change in miRNA levels. We shall therefore term this the `susceptible' regime for ceRNA $i$.
\item[{\bf Repressed regime}~:]When miRNA levels become sufficiently large, ceRNA $i$ will eventually become fully repressed. In order for this to occur, it suffices that $\sum_a \mu_a/\mu_{0,ia}\gg 1$ (which occurs e.g. when the level of at least one of the miRNA species targeting $i$ significantly exceeds its corresponding threshold $\mu_{0,ia}$). We shall call this the `repressed' regime for ceRNA $i$.
\end{description}
(Notice that, because the role of miRNAs and ceRNAs is fully interchangeable, similar regimes can be defined for miRNAs, with the reference concentrations $m_{0,ia}$ playing the role of the threshold ceRNA levels characterizing the distinct regimes.)
\begin{figure}[t]
\begin{center}
\includegraphics[width=16cm]{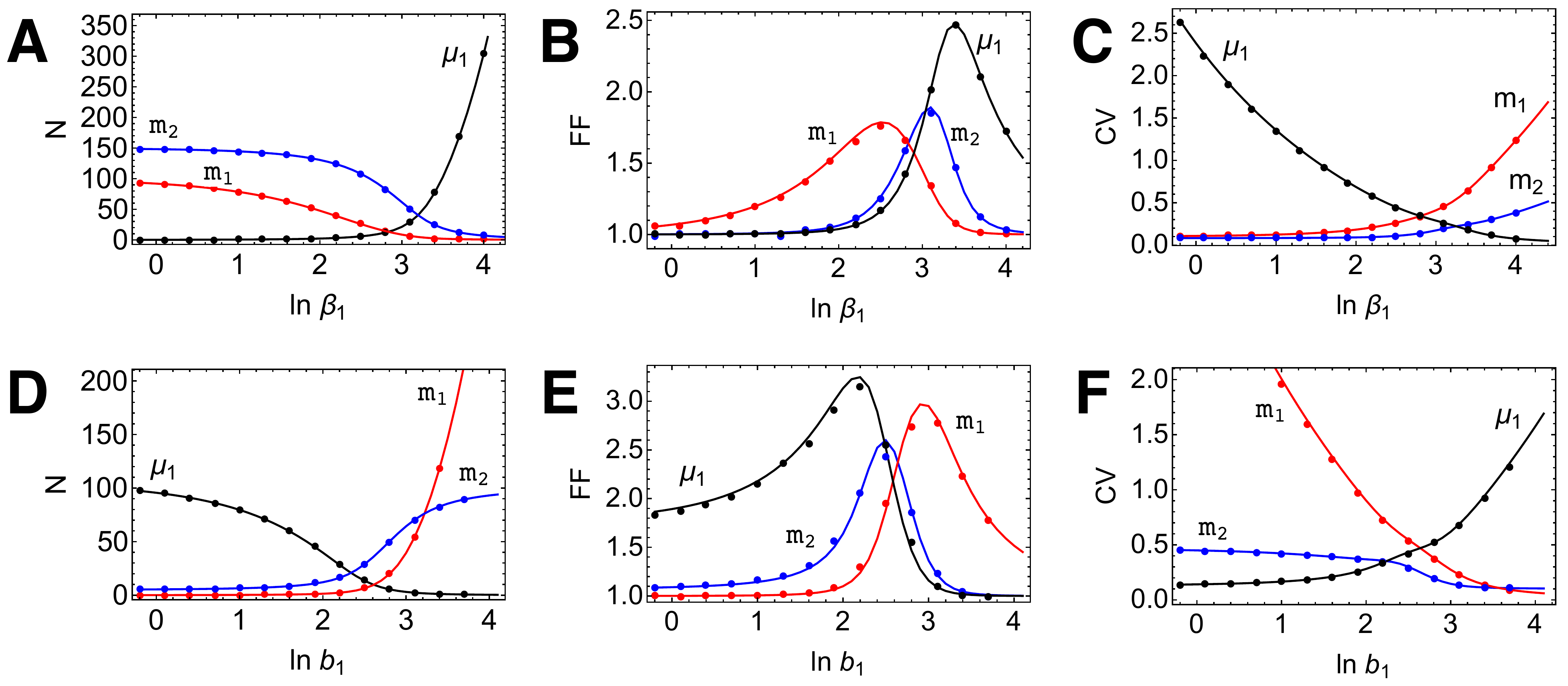}
\end{center}
\caption{Characterization of the steady state for a system with 2 ceRNA species competing for one miRNA species. {\bf (A)} Steady state molecular levels as a function of the miRNA transcription rate $\beta_1$. {\bf (B)} Fano Factor (FF) of each molecular species versus $\beta_1$. {\bf (C)} Coefficient of variation (CV) of each molecular species versus $\beta_1$. {\bf (D)} Steady state molecular levels as a function of the transcription rate of ceRNA 1, $b_1$. {\bf (E)} Fano Factor of each molecular species versus $b_1$. {\bf (F)} Coefficient of variation of each molecular species versus $b_1$. In panels (A) and (D), continuous lines describe analytical results (from Eq~(\ref{ss})) while markers denote mean values obtained from stochastic simulations performed using the Gillespie algorithm (see Sec. \ref{gillespie}). In panels (B), (C), (E) and (F), continuous lines describe analytical results obtained by the Linear Noise Approximation (see Sec. \ref{lna}) while markers represent numerical results derived from stochastic simulations. Parameter values are reported in Table \ref{pars}.}
\label{Fig2}       % Give a unique label
\end{figure}
Fig. \ref{Fig2}A and D report results obtained for the case $N=2$, $M=1$ (two ceRNA species competing for a single miRNA regulator). One sees that ceRNA levels get increasingly repressed as the miRNA transcription rate increases while all other parameters remain fixed (Fig. \ref{Fig2}A). The range of values of $\beta_1$ where ceRNA levels change most strongly corresponds to the susceptible regime. One also sees that ceRNAs 1 and 2 have slightly different thresholds ($\mu_{0,11}\simeq 2$ and $\mu_{0,21}\simeq 15$), as ceRNA 1 is clearly sensitive to variations in miRNA availability for smaller values of $\beta_1$ compared to ceRNA 2. Fig. \ref{Fig2}D shows instead how molecular levels change upon modulating the transcription rate of ceRNA species 1. As $b_1$ increases, $m_1$ grows as expected while concentration of free miRNAs decreases as they increasingly engage targets. This in turn derepresses the other ceRNA species, whose level also increases as the transcription rate of ceRNA 1 is upregulated. That the level of ceRNA 2 can increase upon changing $b_1$ is the key signature of the miRNA-mediated crosstalk that can be established between competing RNAs.
\begin{table}[b]
\caption{Values of kinetic parameters used in the different figures. \label{pars}}
\begin{tabular}{p{2.6cm} p{1.4cm} p{0.9cm} p{0.9cm} p{0.9cm} p{0.9cm} p{0.9cm} p{1.1cm} p{1.1cm} p{1.1cm}}
\hline\noalign{\smallskip}
Parameter & Fig. 2A--C & 2D--F & 3A & 3B & 3C & 3D & 5 & 6A,B & 6C,D\\
\noalign{\smallskip}\svhline\noalign{\smallskip}
$b_1$ [molec min$^{-1}$] & 10 & -- & 10 & 20 & 2 & 1 & 1 (mean) & -- & 10\\
$b_2$ [molec min$^{-1}$] & 15 & 10 & 15 & 10 & 10 & 10 & 1 (mean) & 0 & 0\\
$\beta_1$ [molec min$^{-1}$] & -- & 20 & 15 & -- & 15 & -- & -- & 15 & 15\\
$d_1$ [min$^{-1}$] & 0.1 & 0.1 & 0.1 & 0.1 & 0.1 & 0.1 & 0.005 & 0.1 & 0.1\\
$d_2$ [min$^{-1}$] & 0.1 & 0.1 & 0.1 & 0.1 & 0.1 & 0.1 & 0.005 & 0 & 0.1\\
$\delta_1$ [min$^{-1}$] & 0.1 & 0.1 & 0.1 & 0.1 & 0.1 & 0.1 & 0.01 & 0.1 & 0.1 \\
$k_{11}^+$ [molec$^{-1}$ min$^{-1}$] & $e^{-3}$ & $e^{-2}$ & 1 & $e^{-2}$ & $e^{15}$ & $e^{15}$ & $e^{-2}$ & shown & caption\\
$k_{21}^+$ [molec$^{-1}$ min$^{-1}$] & $e^{-5}$ & $e^{-4}$ & $e^{-3}$ & $e^{-3}$ & $e^{-4}$ & $e^{-4}$ & $e^{-3}$ & 0 & caption\\
$k_{11}^-$ [min$^{-1}$] & 0.001 & 0.001 & 0.001 & 0.001 & 0.001 & 0.001 & 0.1 & 0.001 & 0.001\\
$k_{21}^-$ [min$^{-1}$] & 0.001 & 0.001 & 0.001 & 0.001 & 0.001 & 0.001 & 0.1 & 0 & 0.001\\
$\kappa_{11}$ [min$^{-1}$] & 0.001 & 0.001 & 0.001 & 0.001 & 0.1 & 0.1 & 0.05 & 0.001 & 0.001\\
$\kappa_{21}$ [min$^{-1}$] & 0.001 & 0.001 & 0.001 & 0.001 & 0.001 & 0.001 & 0.05 & 0 & 0.001\\
$\sigma_{11}$ [min$^{-1}$] & 1 & 1 & 1 & 1 & 1 & 1 & 0.001 & 1 & 1\\
$\sigma_{21}$ [min$^{-1}$] & 1 & 1 & 1 & 1 & 1 & 1 & 0.001 & 0 & 1\\
\noalign{\smallskip}\hline\noalign{\smallskip}
\end{tabular}
\end{table}

\paragraph{{\bf \textsf{Note}} }

The reference levels (\ref{thresholds}) ultimately represent the combinations of parameters that are most relevant in order to elucidate many of the network's features. As one would expect, the leading behaviour for $\mu_{0,ia}$ is determined by the ratio $d_i/k_{ia}^+$: the threshold gets smaller as the miRNA-ceRNA interaction gets stronger (i.e., lower miRNA levels suffice to repress a target in presence of stronger coupling), whereas larger intrinsic ceRNA decay rates impose larger repression thresholds. Expectedly, catalytic decay rate affects the thresholds $\mu_{0,ia}$ and $m_{0,ia}$ differentially: while the former decreases as catalytic processing gets more efficient (i.e., miRNA recycling strengthens repression by effectively increasing miRNA availability), $m_{0,ia}$ increases as $\kappa_{ia}$ gets larger (i.e., higher ceRNA levels are required to repress miRNAs at high catalytic processing rates). Note however that $m_{0,ia}$ diverges as $\sigma_{ia}\to 0$, i.e. when all miRNAs are recycled after complex degradation. In other words, in absence of stoichiometric processing of the $i-a$ complex, miRNA $a$ can never be repressed by ceRNA $i$. This implies that, in order for the ceRNA scenario described above to take place, it is necessary that the stoichiometricity ratio $\alpha_{ia}$, Eq. (\ref{sr}), is strictly positive.

\subsection{Stochastic model}
\label{stochastic}

Like all regulatory processes \cite{kond}, the individual reactions reported in (\ref{processes}), i.e. transcription, degradation and titration events due to miRNA-ceRNA interactions, are intrinsically stochastic. This means in practice that molecular levels evolving in time according to (\ref{processes}) are bound to be subject to random fluctuations, with the strength of the noise affecting each molecular species roughly proportional to the square root of its mean. After a transient, concentrations will stabilize and fluctuate around the steady state of the deterministic model (\ref{uno}), described by (\ref{eq:steadystateM}). The deterministic model thereby yields a description of the miRNA-ceRNA network that is all the more accurate when the system is well mixed and concentrations are sufficiently large, making noise negligible. Besides giving a more realistic description of the dynamics of molecular populations, accounting for randomness is however crucial to characterize ceRNA crosstalk in detail, and particularly to disentangle competition-induced effects from fluctuation-induced ones. We shall now therefore briefly review some of the frameworks that have been employed to analyze the stochastic dynamics of (\ref{processes}).

%  beta -> k_S
%  b -> k_R
%  delta -> g_S
%  d -> g_R
%  k^+ -> k^+
%  k^- -> k^-
%  sigma -> alpha * gamma
%  kappa -> (1-alpha) * gamma
%  Da queste ottengo alpha = sigma/(sigma + kappa) quindi il limite
%  catalitico alpha = 0 (oppure di completo riciclo del mirna)
%  equivale al limite sigma -> 0

\subsubsection{The master equation}

The direct mathematical route to account for stochasticity is based on the chemical Master Equation (ME) \cite{vkam}, which describes the time evolution of the probability $P(\pmb{\mu},{\bf m},{\bf c} ,t)$ to find the system with prescribed molecular levels ${\bf m}=\{m_i\}_{i\in\{1,\dots,N\}}$ for ceRNAs, $\pmb{\mu}=\{\mu_a\}_{a\in\{1,\dots,M\}}$ for miRNAs and ${\bf c}=\{c_{\ell}\}_{\ell\in{1,\dots,M\cdot N}}$ for the $N\cdot M$ species of miRNA-ceRNA complexes at time $t$. The ME reads
\begin{align}
\label{eq:masterCompleta}
  &\frac{\partial P}{\partial t} =\sum_{a=1}^{M}\beta_{a}\left(P_{\mu_{a}-1}-P\right)& \, &\emptyset \xrightharpoonup[\beta_a]{}\mu_a&\nonumber\\
&+\sum_{i=1}^{N}b_{i}\left(P_{m_{i}-1}-P\right)& \, &\emptyset\xrightharpoonup[b_i]{}m_i&\nonumber\\
&+\sum_{a=1}^{M}\delta_{a}\left[(\mu_{a}+1)P_{\mu_{a}+1}-\mu_{a}P\right]&\, &\mu_a\xrightharpoonup[\delta_a]{}\emptyset&\nonumber\\
&+\sum_{i=1}^{N}d_{i}\left[(m_{i}+1)P_{m_{i}+1}-m_{i}P\right]&\,  &m_i \xrightharpoonup[d_i]{}\emptyset&\nonumber\\
&+\sum_{i=1}^{N}\sum_{a=1}^{M}k_{ia}^{+}\left[(\mu_{a}+1)(m_{i}+1)P_{\mu_{a}+1,m_{i}+1,c_{ia}-1}-\mu_{a}m_{i}P\right]&\, &\mu_a+m_i\xrightharpoonup[k_{ia}^+]{}c_{ia}&\\
&+\sum_{i=1}^{N}\sum_{a=1}^{M}k_{ia}^{-}\left[(c_{ia}+1)P_{\mu_{a}-1,m_{i}-1,c_{ia}+1}-c_{ia}P\right]& \,  &c_{ia}\xrightharpoonup[k_{ia}^-]{}\mu_a+m_i&\nonumber \nonumber\\
&+\sum_{i=1}^{N}\sum_{a=1}^{M}\sigma_{ia}\left[(c_{ia}+1)P_{c_{ia}+1}-c_{ia}P\right]& \,  &c_{ia}\xrightharpoonup[\sigma_{ia}]{}\emptyset&\nonumber\\
&+\sum_{i=1}^{N}\sum_{a=1}^M\sum_{i=M+1}^{M+N}\kappa_{ia} \left[(c_{ia}+1)P_{\mu_{a}-1,c_{ia}+1}-c_{ia}P\right]& \,  &c_{ia}\xrightharpoonup[\kappa_{ia}]{}\mu_a&\nonumber
\end{align}
where we adopted for simplicity the compact notation $P_{x_i\pm1}:=P(x_1,\dots,x_i\pm1,\dots,x_{N+M+NM})$. Eq~(\ref{eq:masterCompleta}) relies on the (unrealistic) hypothesis that chemical species live in a well mixed environment without compartments, so that they are all in principle capable of interacting. An interesting and fundamental connection between the mass action kinetics in Eq~(\ref{uno}) and the ME is provided by the so-called {\em mean field approximation}, which assumes a simplified factorized form for the joint probability distribution $P$: 
\begin{equation}
\label{eq:meanfield}
P(\{\mu_a\},\{m_i\},\{c_{ia}\},t) =\prod_{i=1}^N P_i(m_i)\prod_{a=1}^M P_a(\mu_a) \prod_{\ell = 1}^{N \cdot M} P_\ell(c_{\ell})
\end{equation}
Plugging (\ref{eq:meanfield}) into (\ref{eq:masterCompleta}) and computing the mean value of all chemical species, one can see that the differential equation governing their the time evolution coincides with Eq.~(\ref{uno}). This point of view casts in a new perspective the deterministic mass action kinetics: as long as the correlations between the different variables can be neglected, the deterministic scheme is expected to provide an accurate description of the dynamics of the model. On the other hand, by construction, the deterministic mass action kinetic is blind to statistical correlations between variables. If one is interested in this aspect, Eq.~(\ref{eq:masterCompleta}) provides the correct theoretical framework.

Unfortunately, the ME is notoriously hard to handle analytically. Therefore, in the following, we will outline different approximation schemes that have been used to obtain useful indications about fluctuations and correlations between molecular levels.

\subsubsection{Gaussian Approximation} 

The Gaussian approximation is probably the simplest one going beyond mean-field. The rationale of the method is rooted in Van Kampen's expansion \cite{vkam}, and specifically in the fact that, if molecules are assumed to be enclosed in a sufficiently large volume, the solution of the ME is Gaussian except for small corrections. Adopting the following vector notation already implicitly used in Eq.~(\ref{eq:meanfield}), i.e.
\begin{eqnarray}
  \vec x &:=& \{x_1 ,\dots x_M ,x_{M+1}\dots x_{M+N},x_{M+N+1},\dots x_{M+N+MN}\} \nonumber\\
  &=& \{\mu_1,\dots,\mu_M,m_1,\dots,m_N,c_{11},\dots,c_{NM}\}\,~~,
\end{eqnarray}
the Gaussian approximation assumes that $\vec x$ is distributed as a multivariate Gaussian, namely
\begin{equation}
\label{eq:gauss-multi}
P(\vec x) \simeq G(\vec x | \vec a, \Sigma^{-1}) = \frac{\exp\left[ -\frac12 \left(\vec x - \vec a \right)^T\Sigma^{-1} \left(\vec x - \vec a \right)\right]}{\sqrt{(2\pi)^{M+N+MN}\mathrm{det}(\Sigma)}}
\end{equation}
where the covariance matrix $\boldsymbol{\Sigma}$ has element $\Sigma_{ij}=E(x_i x_j)-E(x_i)E(x_j)$,  the vector $\vec a$ has coordinates $a_i = E(x_i)$, and the expectation value $E(\cdot)$ is with respect to the Gaussian measure $G$ defined in Eq~(\ref{eq:gauss-multi}). One of the characteristics that make Gaussian distributions useful in this context lies the property that all moments of a Gaussian measure can be expressed in terms of the mean $\vec a$ and the covariance matrix $\boldsymbol{\Sigma}$, so that, for instance, the generic third and fourth order moments read $E(x_i x_j x_k) = \Sigma_{ij}a_k + \Sigma_{ik}a_j + \Sigma_{jk}a_i$ and $  E(x_i x_j x_k x_l)=\Sigma_{ij}\Sigma_{kl}+\Sigma_{ik}\Sigma_{jl}+\Sigma_{il}\Sigma_{jk}$ respectively. In analogy with the closure of the system of equations in the first moments that the factorization hypothesis in Eq~(\ref{eq:meanfield}) induces, a shrewd use of the moment generating function produces a closed system of equations for $\vec a$ and  $\boldsymbol{\Sigma}$. The natural formalism to impose this moment closure is that of the {\em moment-generating function}, defined as
\begin{equation}
  \label{eq:momgen}
  F({\bf z},t) = \sum_{\bf x}\prod_{i=1}^{N+M+N\cdot M}z_i^{x_i}P({\bf x},t)~~.
\end{equation}
It is simple to show that the time evolution of $F({\bf z},t)$ is ruled the second-order partial differential equation
\begin{equation}
\label{eq:momgenevol}
\partial_t F({\bf z},t) = {\cal H}({\bf z})F({\bf z},t)\quad,
\end{equation}
where, for the miRNA-ceRNA network, the operator ${\cal H}$ is defined as
\begin{eqnarray}
  {\cal H}({\bf z}) &=& \sum_{a=1}^M \beta_a (z_a-1) + \sum_{i=M+1}^{M+N}b_i(z_i-1) \nonumber\\
  &+& \sum_{a=1}^M \delta_a (\partial_{z_a}-z_a\partial_{z_a}) + \sum_{i=M+1}^{M+N}d_i(\partial_{z_i}-z_i\partial_{z_i}) + \sum_{l=N+M+1}^{N+M+N\cdot M}\sigma_{l}(\partial_{z_l}-z_l\partial_{z_l})\nonumber\\
  &+& \sum_{a=1}^M\sum_{i=M+1}^{M+N} k_{ia}^+(z_{ia}\partial^2_{z_i\,z_a} - z_i z_a \partial^2_{z_i\,z_a})
  + \sum_{a=1}^M\sum_{i=M+1}^{M+N} k_{ia}^-(z_i z_a \partial_{z_{ia}} - z_{ia} \partial_{z_{ia}})\nonumber\\
  &+& \sum_{a=1}^M\sum_{i=M+1}^{M+N} \kappa_{ia}( z_i \partial_{z_{ia}}-z_{ia}\partial_{z_{ia}})~~.
\end{eqnarray}
The moment-generating function $F$ owes its name to the following
constitutive property:
\begin{equation}
  \partial^{l_1+l_2+\dots+l_k}_{z^{l_1}_{i_1},z^{l_2}_{i_2},\dots,z^{l_k}_{i_k}}F({\bf z},t)|_{\bf z=1}=\langle x_{i_1}^{l_1}x_{i_2}^{l_2}\cdots x_{i_k}^{l_k}\rangle_{P({\bf x},t)}\quad.
\end{equation}
In other terms, consecutive derivatives of $F$ generate all moments of the distribution $P$. The ME Eq~(\ref{eq:masterCompleta}) allows us to write a hierarchy of equations for the moments. However, it turns out that moments of order $k$ are usually expressed in terms of moments of order $k+1$, not allowing to close the system of equations for the moments. The Gaussian approximation truncates the hierarchy of moment dependencies by expressing third-order cumulants in terms of second-order ones (an approximation that turns out to be correct for Gaussian distributions). Thanks to this moment-closure approximation one ends up with a complete system of $N+M+N\cdot M+ {{N+M+N\cdot M}\choose2}$ equations for the mean molecular levels and all covariances.

\subsubsection{The Langevin approach}

A possibly more intuitive description of the stochastic dynamics is obtained by noting that, under broad conditions \cite{vkam}, one can effectively represent molecular fluctuations by adding specific  noise terms to each of the factors appearing in the kinetic Eqs (\ref{uno}). This leads to a Langevin dynamics given by 
\begin{equation}\label{due}
\arraycolsep=1pt\def\arraystretch{2}
\begin{array}{r@{}l}
\frac{d m_i}{dt}&=b_i-d_i m_i\,\uudl{+\xi_i}\,-\sum_a k_{ia}^+ m_i\mu_a \,\uudl{+\sum_a \xi_{ia}^+}\,+\sum_a k_{ia}^- c_{ia}\,\uudl{+\sum_a\xi_{ia}^-}~~, \\
\frac{d \mu_a}{dt}&=\beta_a-\delta_a \mu_a\,\uudl{+\xi_a}\,-\sum_i k_{ia}^+ m_i\mu_a\,\uudl{+\sum_i \xi_{ia}^+}\, + \sum_{i}(k_{ia}^-+\kappa_{ia})c_{ia}\,\uudl{+\sum_i(\xi_{ia}^-+\xi_{ia}^{\rm cat})}~~,\\
\frac{d c_{ia}}{dt}&= k_{ia}^+ m_i \mu_a\,\uudl{+\xi_{ia}^+}\,-(\sigma_{ia}+\kappa_{ia}+k_{ia}^-)c_{ia}\,\uudl{+(\xi_{ia}^{\rm st}+\xi_{ia}^{\rm cat}+\xi_{ia}^-)}~~,
\end{array}
\end{equation}
where the mutually independent stochastic `forces' associated to each process have been inserted after the corresponding term and  underlined. In specific, 
\begin{itemize}
\item $\xi_i$ and $\xi_a$ represent the intrinsic noise due to random synthesis and degradation events that affect $m_i$ and $\mu_a$, respectively; 
\item $\xi_{ia}^+$ and $\xi_{ia}^-$ model the noise affecting the random association and dissociation of complexes, respectively;
\item $\xi_{ia}^{\rm cat}$ and $\xi_{ia}^{\rm st}$ represent the noise of catalytic and stoichiometric complex processing events, respectively. 
\end{itemize}
Each of these noise terms has zero mean. Correlations are instead given by
\begin{equation}\label{lang_noise}
\arraycolsep=1pt\def\arraystretch{1.5}
\begin{array}{r@{}l}
&{} \avg{\xi_{i}(t)\xi_{i}(t')} =  (b_i+d_i \avg{m_i}) ~ \delta(t-t')~~,\\
&{}\avg{\xi_{a}(t)\xi_{a}(t')} = (\beta_a+\delta_a \avg{\mu_a}) ~ \delta(t-t')~~, \\
&{}\avg{\xi_{ia}^+(t)\xi_{ia}^+(t')} =  k_{ia}^+ \avg{m_i} \avg{\mu_a} ~ \delta(t-t')~~,\\
&{}\avg{\xi_{ia}^-(t)\xi_{ia}^-(t')} =  k_{ia}^- \avg{c_{ia}} ~ \delta(t-t')~~,\\
&{}\avg{\xi_{i}^{\rm cat}(t)\xi_i^{\rm cat}(t')} =  \kappa_{ia} \avg{c_{ia}} ~ \delta(t-t')~~,\\
&{}\avg{\xi_{i}^{\rm st}(t)\xi_{i}^{\rm st}(t')} = \sigma_{ia} \avg{c_{ia}} ~ \delta(t-t')~~,
\end{array}
\end{equation}
where steady state abundances (in angular brackets) are given by the solutions of Eqs (\ref{eq:steadystateM}). The specific form (\ref{lang_noise}), involving steady state vaues, can be derived within the so-called Linear Noise Approximation (LNA, \cite{vkam}),  assuming that stationary molecular levels are sufficiently large \cite{swain}. As we show next, the LNA also provides direct access to the covariances of molecular levels.

\subsubsection{Linear Noise Approximation}
\label{lna}

Denoting by $\mathbf{x}$ the vector of molecular levels of all species involved, i.e. $\mathbf{x}= (\{m_i\},\{\mu_a\},\{c_{ia}\})$, the stochastic dynamics (\ref{due}) can be written in vector notation as
\begin{equation}\label{dueprime}
\frac{d\mathbf{x}}{dt}=\mathbf{f(x)}+\boldsymbol{\xi}~~,
\end{equation}
where the vector function $\mathbf{f}$ accounts for the deterministic terms in (\ref{due}) while the vector noise $\boldsymbol{\xi}$ contains the overall noise affecting each component. The LNA is based on the assumption that, at stationarity, random fluctuations cause $\mathbf{x}$ to deviate from is steady state value $\avg{\mathbf{x}}$ by a quantity $\delta \mathbf{x} =  \mathbf{x} - \avg{\mathbf{x}}$ that is small enough to allow for the linearization of Eq (\ref{dueprime}) around $\avg{\mathbf{x}}$. In such conditions,  $\delta \mathbf{x}$ changes in time as \cite{vkam}
\begin{equation}
\label{SDE}
\frac{d }{d t} \delta \mathbf{x} = \mathbf{S} \delta \mathbf{x} + \boldsymbol{\xi}~~~~~,~~~~~\mathbf{S}=\frac{d\mathbf{f}}{d\mathbf{x}}\bigg|_{\mathbf{x=\avg{x}}}~~,
\end{equation}
where $\mathbf{S}$ is the stability matrix of first-order derivatives evaluated at the steady state. Assuming that $\boldsymbol{\xi}$ is a Gaussian noise with zero mean and  cross-correlations described by a matrix $\boldsymbol{\Gamma}$, i.e. $\avg{\xi_s(t) \xi_{s'} (t')} = \Gamma_{ss'} \delta (t-t')$ (where the indices $s$ and $s'$ range over the components of $\mathbf{x}$), one can show that the covariances of molecular levels at steady state obey \cite{swain} 
\begin{equation}
 \avg{ \delta x_a \delta x_b }    = - \sum_ {i, l, s, r } B_{as} B_{br} \frac{\Gamma_{il} } {\lambda_s + \lambda_r }   (B^{-1})_{si} (B^{-1})_{rl} ~~,
\end{equation}
where $\boldsymbol{\lambda}$ denotes the vector of eigenvalues of the stability matrix, while $\mathbf{B}$ stands for its eigenvectors (i.e. $\sum_{b}S_{ab}B_{br}=\lambda_r B_{ar}$).

The above formula provides a way to estimate correlations (and hence Pearson coefficients) of all molecular species involved in the system. The continuous lines in Fig. \ref{Fig2}B, C, E and F have indeed been obtained by the LNA.

\subsubsection{The Gillespie algorithm}
\label{gillespie}

The standard numerical route to simulate systems like Eq (\ref{due}) relies on the Gillespie algorithm (GA), a classical stochastic simulation method that computes the dynamics of a well-mixed  system of molecular species interacting through a set of possible processes \cite{gill}. The GA allows to simulate the dynamics of systems like (\ref{processes}) without solving the ME, i.e. without the full knowledge of the probability $P(\mathbf{x},t)$ of the system being in state vector $\mathbf{x}$ (encoding for the population of each molecular species) at time $t$. In short (see however \cite{gibs} for a more detailed presentation), one can say that the GA essentially relies on two assumptions: (i) each process occurs with a specific rate constant; and (ii) the current state of the system (in terms of the number of molecules of each species) determines which process is going to occur next, independently of the previous history. Under these conditions, one can simulate trajectories of a system described by a set of processes such as (\ref{processes}) simply from the knowledge of the probability density $P(k,\tau|\mathbf{x},t)$ that process $k$ takes place between time points $t+\tau$ and $t+\tau+d\tau$ given that the state of the system at time $t$ is $\mathbf{x}$ (with no other processes occurring between time  $t$ and time $t+\tau$). Because the dynamics is memoryless, $P(k,\tau|\mathbf{x},t)$ factorizes as
\begin{equation}\label{gilles}
\arraycolsep=1pt\def\arraystretch{1.5}
\begin{array}{r@{}l}
P(k,\tau|\mathbf{x},t)d\tau=&\,\,\text{Prob}\{\text{no process between time $t$ and time $t+\tau$}\}\times\\
& \,\,\times \,\,\text{Prob}\{\text{process $k$ between time $t+\tau$ and time $t+\tau+d\tau$}\}\\
\equiv& \,\, P_0\times P_k~~.
\end{array}
\end{equation}

The probability $P_k$ is given by the intrinsic rate of process $k$ ($c_k$) times a function of $\mathbf{x}$ ($g_k(\mathbf{x})$) that quantifies the number of different ways in which process $k$ might occur and which basically encodes for the law of mass action. We shall use the shorthand $c_k g_k(\mathbf{x})=f_k(\mathbf{x})$. Hence $P_k=f_k(\mathbf{x}(t+\tau))d\tau$.

$P_0$ can instead be evaluated by sub-dividing the interval $[t,t+\tau]$ in $K$ parts ($K\gg 1$), each of duration $\tau/K$. If $f_k$ denotes the rate of process $k$, then $P_0$ is just the probability that no process occurs in any of the $K$ sub-intervals, i.e.
\begin{equation}
P_0 = \left(1-\sum_{k'} f_{k'}\frac{\tau}{K}\right)^K\simeq e^{-\tau \sum_{k'} f_{k'}}~~~~~(K\gg 1)~~.
\end{equation}
Hence
\begin{equation}\label{gilles2}
P(k,\tau|\mathbf{x},t)\simeq f_k \,\, e^{-\tau \sum_{k'} f_{k'}}~~,
\end{equation}
which can also be re-cast as
\begin{equation}\label{gilles3}
P(k,\tau|\mathbf{x},t)\simeq \underbrace{\left(\sum_{k'}f_{k'}\right) e^{-\tau \sum_{k'} f_{k'}}}_{\text{prob. of waiting time $\tau$}} \quad\times\,\underbrace{\frac{f_k}{\sum_{k'}f_{k'}}}_{\text{prob. of process $k$}} \,\, ~~.
\end{equation}
A value of $\tau$ sampled from the above distribution of waiting times is easily obtained by noting that, if $u$ denotes a random variable uniformly distributed in $[0,1]$, then 
\begin{equation}
\tau=-\frac{\ln(u)}{\sum_{k'}f_{k'}}
\end{equation}
is actually distributed according to the exponential function given in (\ref{gilles3}). This allows to formulate the GA in the following scheme:
\begin{svgraybox}
{\bf \textsf{Gillespie Algorithm} }
\begin{description}  
\setlength\itemsep{0pt}
\item[Step 1:] Initialization: set initial populations for all molecular species (vector $\mathbf{x}(0)$) together with the rate $c_k$ of each process $k$ and an end-time $T$
\item[Step 2:] Evaluate reaction probabilities $f_k$ for each $k$ as well as $\sum_{k'}f_{k'}\equiv Z$
\item[Step 3:] Generate a pair $(k,\tau)$ from (\ref{gilles3})
\item[Step 4:] Update molecular populations according to the selected process $k$ and advance time by $\tau$
\item[Step 5:] Iterate from Step 2 or stop if the end-time $T$ has been reached
\end{description}
\end{svgraybox}

Fig. \ref{Fig2}B, C, E and F show how mean molecular levels obtained by the GA (markers) compare against analytic results (lines). One sees that the Fano Factor (FF) markedly peaks when molecular levels become roughly equimolar, i.e. close to the threshold where the system becomes susceptible to changes in the modulated parameter (in this case, the miRNA transcription rate or the transcription rate of ceRNA 1). The coefficient of variation (CV) also modifies its qualitative behaviour in the same range, although this feature generically appears to be less drastic (see however \cite{bosi}). This shows that when ceRNAs become susceptible and cross-talk is established, fluctuations in molecular levels become strongly correlated. 

The fluctuation scenario just described is clearly connected to the establishment of miRNA-mediated crosstalk. How exactly, and how it relates to other signatures of cross-talk, is the subject of the following section.

\subsection{Quantifying miRNA-mediated crosstalk at steady state}
\label{subsec:3}

The competing endogenous RNA scenario concerns the possibility that, as a result of competition to bind miRNAs, ceRNAs could cross-regulate each other. We have so far identified two  signatures that accompany the establishment of miRNA-mediated crosstalk at stationarity: 
\begin{enumerate}
\item[a.] a change in the steady state level of a ceRNA following a change of the level of a competitor (i.e. a response following a perturbation); 
\item[b.] an increase of connected ceRNA-ceRNA correlations. 
\end{enumerate}
Both are clearly defined and  testable in experiments and from data (at least in principle). Yet, despite the apparent simplicity, the reliable detection of the ceRNA mechanism in experiments or data is far from simple. The key issue lies in the fact that several mechanisms, both involving miRNAs and involving other molecular actors, potentially bear similar effects on transcripts and, as the cause differs, so do their consequences. Disentangling the  competition-driven ceRNA effect from other processes is in many ways essential to be able to predict how a miRNA-ceRNA network will react to perturbations. We shall recap below how the ceRNA crosstalk scenario looks when seen through different glasses. While each allows to capture certain aspects of the ceRNA mechanism, different quantities employed to quantify crosstalk  intensity focus on slightly different physical features and therefore can be useful in different situations.  Understanding such differences is however crucial both for applications and for the unambiguous identification of biological drivers.

\subsubsection{Pearson correlation coefficient}

Since an increase of correlations between molecular levels accompanies the establishment of crosstalk, it is reasonable to view the Pearson correlation coefficient between two ceRNAs as a basic proxy for crosstalk intensity \cite{alau,bosi,meht}. For ceRNAs $i$ and $j$, it is defined as
\begin{equation}\label{rho}
\rho_{ij}=%\frac{\avg{m_i m_j}_c}{\sigma_{m_i}\sigma_{m_j}}\equiv
\frac{\avg{m_i \,m_j}-\avg{m_i}\avg{m_j}}{\sqrt{\avg{m_i^2}-\avg{m_i}^2}\sqrt{\avg{m_j^2}-\avg{m_j}^2)}}\equiv\frac{{\rm cov}(m_i,m_j)}{\sqrt{\avg{\delta m_i^2}}\,\sqrt{\avg{\delta m_j^2}}}~~,
\end{equation}
where averages are taken over random fluctuations in the steady state of a stochastic dynamics. (When the interaction network is conserved across different cellular samples and single snapshots of molecular levels are available for each sample, the $\avg{\cdots}$ average can also be taken over different samples, as long as each sample can be considered to be stationary.) Note that $-1\leq\rho_{ij}\leq 1$. 

The rationale for using (\ref{rho}) as a measure of crosstalk intensity is roughly the following. In a network of $N$ ceRNA species interacting with $M$ miRNA species, both ceRNA and miRNA levels will fluctuate stochastically over time at stationarity. A large positive value of $\rho_{ij}$ points to the existence of a positive (linear) correlation between $m_i$ and $m_j$, i.e. to the fact that $m_i\simeq c m_j+d+$noise, with constants $c>0$ and $d$. In such conditions, it is reasonable to expect that an increase in the level of ceRNA $i$, whichever its origin, will divert part of the miRNA population currently targeting ceRNA $j$ to bind to $i$, thereby freeing up molecules of $j$ for translation. In practice, with a large $\rho_{ij}$, perturbations affecting ceRNA $i$ could be ``broadcast'' to ceRNA $j$ because of the miRNA-mediated statistical correlation existing between their respective levels.

\begin{figure}[t]
\begin{center}
\includegraphics[width=16cm]{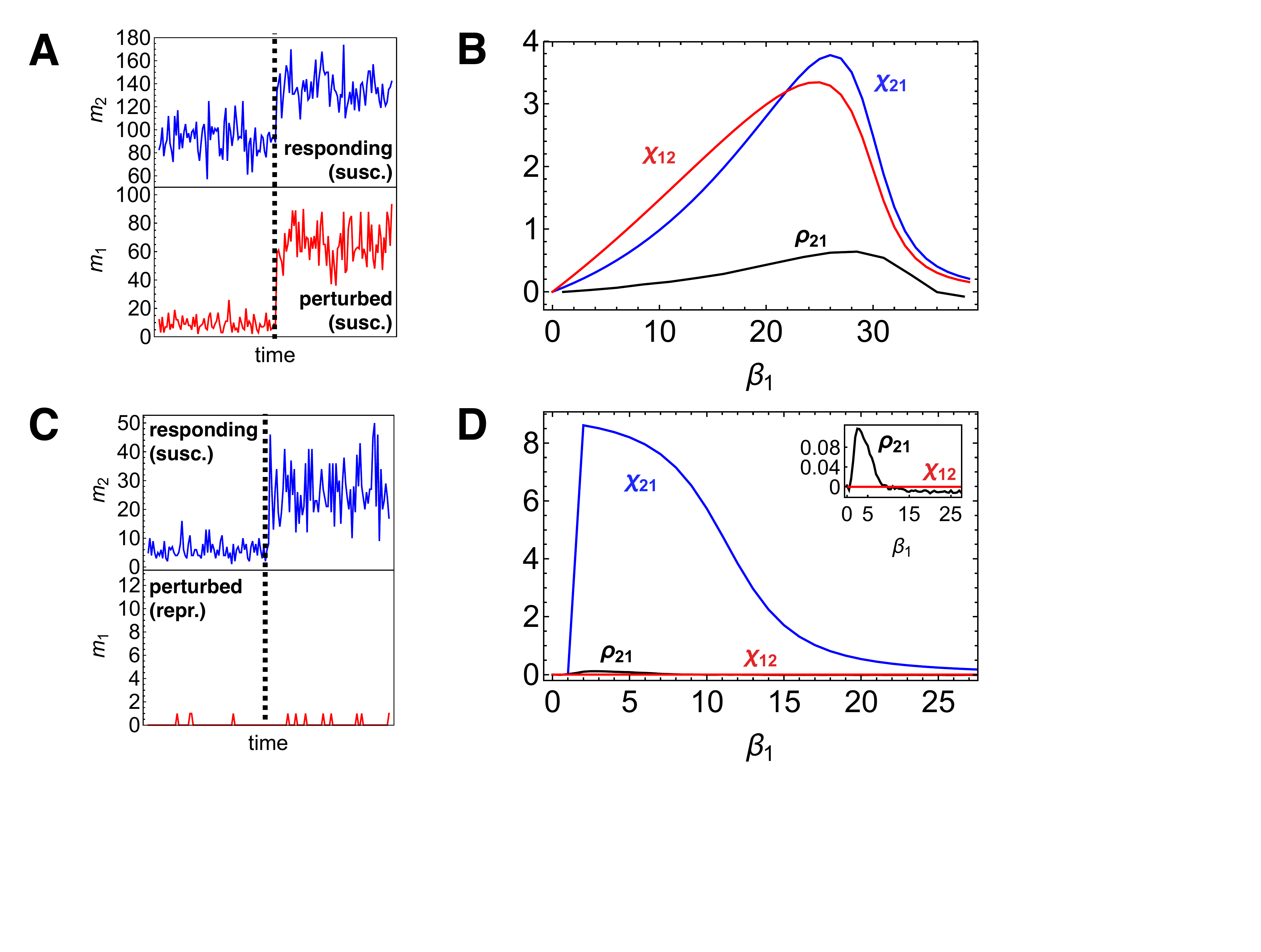}
\end{center}
\caption{{\bf (A)} Stochastic simulation showing the free levels of two ceRNA species co-regulated by a miRNA species (not shown). Both ceRNAs are susceptible with respect to changes in the miRNA level. The transcription rate of ceRNA 1 is perturbed at the time indicated by the dashed line. ceRNA 2 responds by increasing its amount. {\bf (B)} Susceptibilities and Pearson coefficients for two ceRNAs co-regulated by a miRNA species for moderate miRNA repression strength. All three quantifiers of ceRNA crosstalk are significantly different from zero and $\chi_{12}\simeq \chi_{21}$. {\bf (C)} Same as (A) but now ceRNA 1 is fully repressed by the miRNA. Still, an increase of its transcription rate yields an upregulation of $m_2$. {\bf (D)} Same as (B) but for strong miRNA repression on ceRNA 1. Both the Pearson coefficient $\rho_{21}$ and $\chi_{12}$ (quantifying the response of ceRNA 1 to a perturbation affecting ceRNA 2) are effectively zero, whereas $\chi_{21}$ is not. Parameter values are given in Table \ref{pars}.}
\label{Fig3}       % Give a unique label
\end{figure}

The Pearson correlation coefficient between competing ceRNAs  indeed attains a maximum in a specific range of values for the  transcription rates, see e.g. Fig. \ref{Fig3}B. 
Expectedly, this happens when the levels of the different molecular species become comparable (or, more precisely, when the number of miRNA binding sites becomes similar to that of miRNA molecules) \cite{bosi}. Here, ceRNA fluctuations become strongly correlated and one might expect ceRNA crosstalk to be active, so that a perturbation affecting one ceRNA will result in a shift in the level its competitor. In other words, this regime is characterized by significant crosstalk effects.

\subsubsection{Susceptibility}

A mechanistic (as opposed to statistical) quantification of the magnitude of the ceRNA effect can be obtained by computing derivatives of steady-state ceRNA levels like \cite{figl}
\begin{equation}\label{sus}
\arraycolsep=1pt\def\arraystretch{1.9}
\begin{array}{r@{}l}
\chi_{ij} &{}= \frac{\partial \avg{m_i}}{\partial b_j}\geq 0\\
\chi_{ia} &{}= \frac{\partial \avg{m_i}}{\partial \beta_a}\leq 0
\end{array} 
\end{equation} 
where $b_j$ (resp. $\beta_a$) stands for the transcription rate of ceRNA $j$ (resp. miRNA $a$). We shall term quantities like (\ref{sus}) {\it susceptibilities}. In short, $\chi_{ij}$ measures the variation in the mean level of ceRNA $i$ caused by a (small) change in $b_j$. As an increase of $b_j$ leads to an increase of the level of ceRNA $j$ by titration of miRNAs away from it, $\chi_{ij}$ is bound to be non-negative. A similar straightforward interpretation applies to $\chi_{ia}$, which is non-positive since an increase of $\beta_a$ is bound to cause a decrease of $\avg{m_i}$. The central hypothesis behind Eq (\ref{sus}) is that small perturbations cause small changes in molecular levels, or, more precisely, that the latter will be proportional to the former if the perturbation is sufficiently small (linear response scenario).

Assuming no direct control of ceRNA $i$ by ceRNA $j$, a large value of $\chi_{ij}$ directly points to the existence of miRNA-mediated crosstalk in terms of a change in the level of a target upon perturbing the level of a competitor. Hence $\chi_{ij}$ focuses on the response part of the ceRNA effect rather than on the fluctuation-related aspects. 

Quantities like $\chi_{ij}$ can be directly computed from the steady state conditions and in numerical simulations upon probing the system with the desired perturbation. A susceptibility-based theory of ceRNA crosstalk at steady state has indeed been presented in \cite{figl}. When quantified through $\chi_{ij}$, ceRNA crosstalk displays the following key features:
\begin{description}
\item[{\bf Selectivity}~:]When a miRNA targets multiple ceRNA species, crosstalk may occur only among a subset of them. This effect is related to the fact that different ceRNAs can have different thresholds for repression by the miRNA and is enhanced by heterogeneities in the thresholds;
\item[{\bf Directionality (asymmetry)}~:]In general, $\chi_{ij}\neq\chi_{ji}$, i.e. ceRNA $i$ may respond to a perturbation affecting ceRNA $j$ but not the reverse;
\item[{\bf Plasticity}~:]The pattern of miRNA-mediated ceRNA crosstalk, whereby ceRNA $j$ is linked to ceRNA $i$ when $\chi_{ij}$ is sufficiently large, is modulated by kinetic parameters, and particularly by miRNA levels (in other words, changes in miRNA availability modify the ceRNA crosstalk network);
\item[{\bf Dependency on stoichiometric processing}~:]If all miRNA-ceRNA complexes formed by ceRNA $j$ are degraded in a purely catalytic way, then $\chi_{ij}=0$ (i.e. stoichiometric processing is necessary for ceRNA crosstalk at stationarity).
\end{description}

Like the Pearson coefficient $\rho_{ij}$, the ceRNA-ceRNA susceptibility $\chi_{ij}$ also peaks when ceRNA crosstalk is strongest (see Fig. \ref{Fig3}B). However, the fact that susceptibilities are perturbation-specific makes their usefulness for data analysis and the interpretation of experiments less immediate compared to Pearson coefficients. Ideally, one would like to connect susceptibilities like (\ref{sus}) to simpler quantities like correlation functions. A more refined mathematical analysis of the stochastic dynamics shows that this is indeed possible.

\subsubsection{Fluctuations versus response}

It is important to understand that the physical meaning and therefore the crosstalk scenarios underlied by $\rho_{ij}$ and $\chi_{ij}$ are rather different. The fact that $\chi_{ij}$ is asymmetric under exchange of its indices (i.e. $\chi_{ij}\neq\chi_{ji}$ in general) whereas $\rho_{ij}$ is necessarily symmetric already pointed in this direction. Other subtle differences however emerge when the two quantities are compared in greater detail. 

In first place, $\chi_{ij}$ can be non zero (and possibly large) even for a completely deterministic system like (\ref{uno}), as it simply measures how a target's steady state level is modulated by changes affecting the transcription rate of one of its competitors, independently of the presence of stochastic fluctuations around the steady state. In this sense, $\chi_{ij}$ focuses exclusively on the effects induced by competition. On the other hand, in absence of fluctuations $\rho_{ij}$ is identically zero. 

Second, and related to this, is the fact that a large value of $\rho_{ij}$ can occur when both ceRNAs respond to fluctuations in miRNA levels (`indirect correlation'). This however does not imply that $m_i$ and $m_j$ are directly correlated. (If variables $X$ and $Y$ are both correlated with $Z$, they will be correlated too. However, in absence of a direct correlation between $X$ and $Y$, upon conditioning over the value of $Z$ one will observe that $X$ and $Y$ are uncorrelated.) The same holds in presence of extrinsic noise, in which case averages are performed over different samples rather than over time in a single sample. To see this directly, one can consider a system formed by $N$ ceRNA species (labeled $i,j,k,\ldots$) and $M$ miRNA species (labeled $a$) \cite{figl}. If transcription rates fluctuate across cells and if fluctuations are sufficiently small, ceRNA levels at steady state will be approximately given by
\begin{equation}\label{cov}
\arraycolsep=1pt\def\arraystretch{1.9}
\begin{array}{r@{}l}
\avg{m_i} &{}\simeq \ovl{\avg{m_i}}+\sum_{j} \frac{\partial \avg{m_i}}{\partial b_j}(b_j-\ovl{b_j})+\sum_a\frac{\partial {\avg{m_i}}}{\partial \beta_a}(\beta_a-\ovl{\beta_a})\\
&{}\equiv \ovl{\avg{m_i}}+\sum_{j} \chi_{ij}\delta b_j+\sum_a\chi_{ia}\delta\beta_a ~~,
\end{array}
\end{equation}
the over-bar denoting an average over transcription rates. Assuming that transcription rates of different species are mutually independent, the Pearson correlation coefficient $\rho_{ij}$ can be seen to be given by
\begin{equation} 
\label{pearson1}
\rho_{ij}=A\,\left(\sum_k \chi_{ik}\chi_{jk}\ovl{\delta b_k^2}+\sum_a \chi_{ia}\chi_{ja}\ovl{\delta \beta_a^2}\right)
 \end{equation}
where $A>0$ is a constant, the index $k$ runs over ceRNAs, the index $a$ runs over miRNAs and $\ovl{\delta b_k^2}$ (resp. $\ovl{\delta \beta_a^2}$) is the variance of the transcription rate of ceRNA species $k$ (resp. miRNA species $a$). Now one sees that, if all ceRNA-ceRNA susceptibilities are zero (i.e. in absence of competition-induced crosstalk), 
\begin{equation} 
 \rho_{ij} \propto  \sum_a \chi_{ia}\chi_{ja}\ovl{\delta \beta_a^2}~~. 
 \end{equation} 
Because ceRNAs always respond to fluctuations in miRNA levels, susceptibilities on the right-hand side are not zero. In particular, both $\chi_{ia}$ and $\chi_{ja}$ are negative, as an increase in miRNA levels causes a decrease in the level of free ceRNAs. One therefore concludes that $\rho_{ij}>0$ even though all ceRNA-ceRNA susceptibilities are nil. This explicitly shows that $\chi_{ij}$ and $\rho_{ij}$ describe {\it a priori} different crosstalk mechanisms.

A mathematical analysis of susceptibilities and fluctuations shows that crosstalk intensity ultimately depends on whether the involved ceRNAs are unrepressed, susceptible or repressed by miRNAs. In particular, it turns out that the ceRNA-ceRNA susceptibility $\chi_{ij}$ is qualitatively described by a matrix whose entries depend only on the state of repression of $i$ (the responding ceRNA) and $j$ (the perturbed one), given by \cite{figl}
\begin{equation}
\chi_{ij}\,\,=\,\,
\begin{tabular}{c c|c|c|c|}
 
  \multicolumn{2}{c|}{~} & \multicolumn{3}{c|}{$j$ (perturbed)}  \\ 
 \multicolumn{2}{c|}{~} & \multicolumn{1}{c|}{Unrepr.} & \multicolumn{1}{c|}{Susc.} & \multicolumn{1}{c|}{Repr.} \\
 \hline
 \parbox[t]{3mm}{\multirow{3}{*}{\rotatebox[origin=c]{90}{$i$ (resp.)}}} & Unrepr. & $\simeq 0$ & $\simeq 0$ & $\simeq 0$\\
 & Susc. & $\simeq 0$ & \cellcolor{gray!50}{$>0$} & \cellcolor{gray!50}{$>0$}\\
 & Repr. & $\simeq 0$ & $\simeq 0$ & $\simeq 0$\\
 \hline
 \end{tabular}~~.
\end{equation}
Besides showing explicitly that $\chi_{ij}\neq \chi_{ji}$, the above matrix clarifies that a non-zero $\chi_{ij}$ (and therefore competition-driven response of $i$ to a change in the transcription rate of $j$) occurs (i) symmetrically, when both ceRNAs are susceptible to the miRNA (as in Fig. \ref{Fig3}A), and (ii) asymmetrically, when the perturbed ceRNA is repressed while the responding one is susceptible (as in Fig. \ref{Fig3}C). Along the same lines, one finds that \cite{tran}
\begin{equation}
\rho_{ij}\,\,=\,\,
\begin{tabular}{c c|c|c|c|}
 
  \multicolumn{2}{c|}{~} & \multicolumn{3}{c|}{$j$ (perturbed)}  \\ 
 \multicolumn{2}{c|}{~} & \multicolumn{1}{c|}{Unrepr.} & \multicolumn{1}{c|}{Susc.} & \multicolumn{1}{c|}{Repr.} \\
 \hline
 \parbox[t]{3mm}{\multirow{3}{*}{\rotatebox[origin=c]{90}{$i$ (resp.)}}} & Unrepr. & $\simeq 0$ & $\simeq 0$ & $\simeq 0$\\
 & Susc. & $\simeq 0$ & \cellcolor{gray!50}{$>0$} & $\simeq 0$\\
 & Repr. & $\simeq 0$ & $\simeq 0$ & $\simeq 0$\\
 \hline
 \end{tabular}~~.
\end{equation}
i.e. the Pearson coefficient should expected to be significantly different from zero only when both ceRNAs are susceptible to changes in miRNA levels, as is clear by comparing Figures \ref{Fig3}B and D. 

The quantitative relationship linking susceptibilities to fluctuations emerges through a more careful mathematical analysis  of Eq~(\ref{due}) based on approximating the stochastic variability affecting molecular levels with a thermal-like noise. This leads to a set of results closely related to the Fluctuation-Dissipation Relations that characterize the linear-response regime of multi-particle systems in statistical physics. Specifically, one finds that, under broad conditions, susceptibilities can be expressed in terms of covariances of molecular levels or functions thereof. In particular,  in Ref. \cite{tran} it is shown that
\begin{equation}\label{covar}
\arraycolsep=1pt\def\arraystretch{1.9}
\begin{array}{r@{}l}
%\avg{m_i}=-T\frac{\partial}{\partial d_i}\log Z(T)~~,\\
\chi_{ij} &{}\equiv\frac{\partial \avg{m_i}}{\partial b_j}=\gamma\,\,  {\rm cov}(m_i,\log m_j)\geq 0~~,\\
\omega_{ij} &{}\equiv\frac{\partial\avg{m_i}}{\partial d_j}=-\gamma\,\,{\rm cov}(m_i,m_j)\leq 0~~,\\
\chi_{ia} &{}\equiv\frac{\partial \avg{m_i}}{\partial \beta_a}=\gamma\,\,  {\rm cov}(m_i,\log \mu_a)\leq 0~~,\\
\omega_{ia} &{}\equiv\frac{\partial\avg{m_i}}{\partial \delta_a}=-\gamma\,\,{\rm cov}(m_i,\mu_a)\geq 0~~,\\
\end{array}
\end{equation}
where $\gamma>0$ is a constant. In other terms, the response $\chi_{ij}$ of $\avg{m_i}$ to a perturbation affecting the transcription rate of ceRNA $j$ is proportional to the covariance function ${\rm cov}(m_i,\log m_j)$ which incidentally, like $\chi_{ij}$, is not symmetric under the exchange of $i$ and $j$. Similarly, the bare ceRNA-ceRNA covariance ${\rm cov}(m_i,m_j)$ describes the response of $\avg{m_i}$ to (small) change of the intrinsic {\it degradation rate} $d_j$ of ceRNA $j$. Importantly, by comparing (\ref{rho}) with $\omega_{ij}$, Eq. (\ref{covar}), one sees that, perhaps unexpectedly, the Pearson coefficient $\rho_{ij}$ is related to $\omega_{ij}$ (rather than $\chi_{ij}$). (Likewise, one could calculate ceRNA-miRNA susceptibilities like $\chi_{ia}$ and $\omega_{ia}$ by evaluating bare covariances of ceRNA and miRNA levels as shown in (\ref{covar}).) 

Generically, covariances are as easy to estimate from transcriptional data as Pearson coefficients, from which they only differ by the (crucial) normalization factor corresponding to the magnitude of fluctuations of individual variables. Relationships (\ref{covar}) have been used to infer different features of ceRNA crosstalk network generated by the tumor suppressor gene PTEN from transcriptional data, in particular directionality \cite{tran}. The large-scale use of such quantities might provide detailed transcriptome-wide crosstalk patterns, open for analysis and further validation.

\subsection{The role of network topology}

The topology of the miRNA-ceRNA provides an additional degree of freedom through which the effectiveness of ceRNA crosstalk can be influenced. To understand how, we assume that the miRNA-ceRNA network is sufficiently sparse and that connectivity correlations are absent. In such conditions, one can reasonably neglect ceRNA-ceRNA couplings involving more than one miRNA species and express the ceRNA-ceRNA susceptibility as
\begin{equation}\label{chiija}
\chi_{ij}\simeq\sum_a\underbrace{\frac{\partial m_i}{\partial \mu_a}\frac{\partial \mu_a}{\partial b_j}}_{\chi_{ij,a}}~~.
\end{equation}
One sees that if $\chi_{ij,a}$, i.e. the ceRNA-ceRNA susceptibility mediated by miRNA species $a$, is roughly the same for all miRNA regulators shared by $i$ and $j$, i.e. if $\chi_{ij,a}\simeq \chi_{ij}^{(0)}$ for all $a$, then $\chi_{ij}\simeq n_{ij}\chi_{ij}^{(0)}$, with $n_{ij}$ the number of miRNA species that target both ceRNAs  $i$ and $j$. In other words, $\chi_{ij}$ increases with the number $n_{ij}$ of miRNA species shared by $i$ and $j$. This dependence can become especially significant in presence of strong degree correlations in the miRNA-ceRNA network, explaining why clustered networks such as those addressed in \cite{bosi} generically lead to more intense crosstalk patterns than random networks.  

The role of topology is however most clearly isolated when ingredients other than strictly topological ones are as homogeneous as possible. We therefore assume that
\begin{enumerate}
\item[a.] all kinetic parameters are homogeneous (i.e. independent of the molecular species); in particular $\mu_{0,ia}\equiv \mu_0$ for all miRNA-ceRNA pairs; 
\item[b.] miRNA levels are homogeneous, that is $\mu_a=\mu$ for each $a$.
\end{enumerate}
Based on these, one can show that, when the number $n_i$ (resp. $n_j$) of miRNAs targeting ceRNA $i$ (resp. ceRNA $j$) is sufficiently large, each shared miRNA contributes a quantity \cite{figl}
\begin{equation}\label{chiija2}
\chi_{ij,a}\simeq \frac{1}{d}\,\,\frac{\til{\mu}}{A+\sum_{k\in a}\frac{1}{1+n_k\til{\mu}}}\,\,\frac{1}{(1+n_i\til{\mu})^2 (1+n_j\til{\mu})}~~
\end{equation}
to the overall susceptibility Eq~(\ref{chiija}), where $\til{\mu}\equiv\mu/\mu_0$ is the miRNA level expressed in units of $\mu_0$, $A>0$ is a constant while $k\in a$ denotes the set of ceRNAs that interact with miRNA $a$. Hence $\chi_{ij,a}$ decreases (i.e. crosstalk intensity is diluted) as $n_i$ increases, as $n_j$ increases, and/or as the number of targets of miRNA $a$ increases. 

This suggests that a particularly intriguing scenario arises when a large number of miRNA species target $i$ and $j$ and when $\mu\ll\mu_0$, i.e. when all ceRNA species are unrepressed by miRNAs. For simplicity, we assume the miRNA-ceRNA interaction network to be a regular bipartite graph where each ceRNA interacts with $n_i=n$ miRNAs while each miRNA interacts with $\nu_\alpha=\nu$ ceRNAs. In this case, (\ref{chiija2}) takes the form 
\begin{equation}\label{chiija3}
\chi_{ij,a}\simeq \frac{1}{d}\,\,\frac{\til{\mu}}{\nu+A(1+n\til{\mu})}\,\,\frac{1}{(1+n\til{\mu})^2}~~.
\end{equation}
Now the value of $\chi_{ij,a}$ clearly depends on $\til{\mu}$. In particular, one sees that
\begin{equation}\label{casi}
\chi_{ij,a}
\begin{cases}
\ll \frac{1}{dn} & \text{for $\mu\ll\mu_0/n$}\\
\simeq \frac{1}{dn} &\text{for $\mu\simeq\mu_0/n$}\\
\ll \frac{1}{dn} & \text{for $\mu\gg\mu_0/n$}\\
\end{cases}~~.
\end{equation}
In other terms, $\chi_{ij,a}$ is maximum when miRNA levels are close to $\mu_0/n$, i.e. (for sufficiently large $n$) when each is well below the susceptibility threshold. 

\begin{figure}[t]
\begin{center}
\includegraphics[width=16cm]{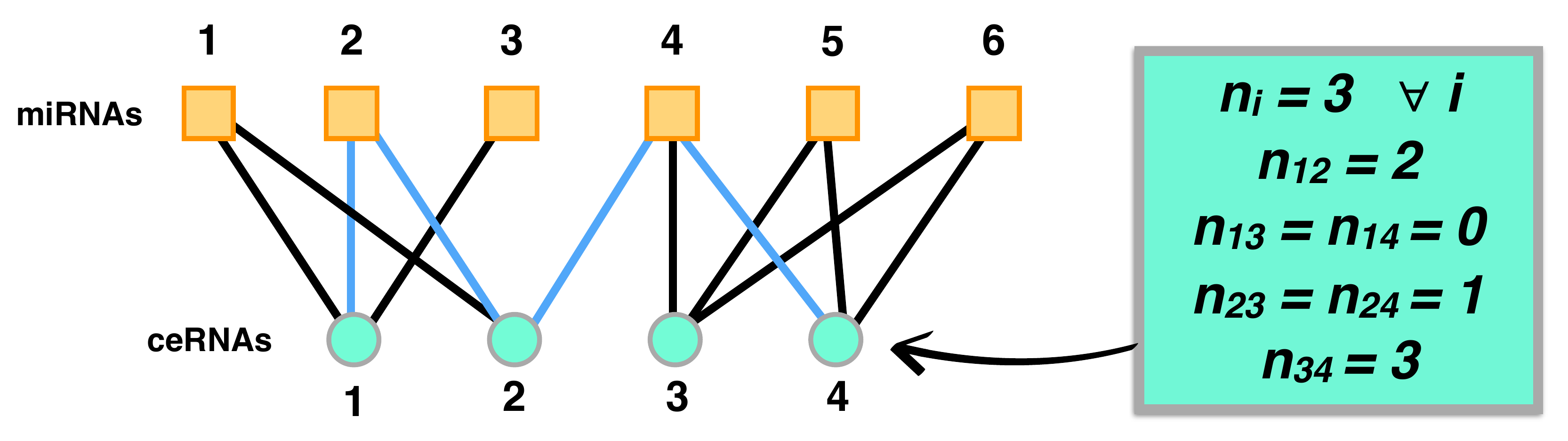}
\end{center}
\caption{Sketch of a miRNA-ceRNA network with $N=4$ and $M=6$. Each ceRNA species is regulated by 3 miRNA species, but ceRNA pairs $(1,2)$ and $(3,4)$ share more regulators than other pairs. Crosstalk between 1 and 2 and between 3 and 4 should therefore generically be stronger than for other ceRNA pairs. On the other hand, ceRNAs 1 and 4 don't have any regulator in common. Still, they may be able to crosstalk through the chain of miRNA-mediated interactions shown in light blue.}
\label{Fig5}       % Give a unique label
\end{figure}

Formula (\ref{casi}) essentially reproduces the standard 3-regime scenario (unrepressed, susceptible, repressed) in a network context, albeit starting from the assumption that ceRNAs are unrepressed by each individual miRNA species. In this sense, it describes a ``distributed'' effect: many weakly interacting miRNA species can collectively mediate efficient ceRNA crosstalk. Recalling (\ref{chiija}), we see that when $\mu\simeq\mu_0/n$ the overall susceptibility is given by
\begin{equation}
\chi_{ij}\propto \frac{n_{ij}}{d n}~~,
\end{equation}
which becomes comparable to the self-susceptibility $\chi_{ii}$ for $n_{ij}\simeq n$. A sketch summarizing the results just described is shown in Fig. \ref{Fig5}.

When connectivity correlations are not negligible and the approximation (\ref{chiija}) fails, $\chi_{ij}$ can in principle be expressed as
\begin{equation}
\chi_{ij}=\sum_{n\geq 0} \chi_{ij}^{(n)}~~,
\end{equation}
where $\chi_{ij}^{(n)}$ stands for the contribution to the $i-j$ susceptibility given by crosstalk interactions mediated by chains formed by $n$ miRNA species. Starting from the steady state conditions (\ref{eq:steadystateM}), one can compute $\chi_{ij}^{(n)}$ exactly in the limit where the stoichiometricity ratio $\alpha_{ia}$ is the same for all pairs, i.e. $\alpha_{ia}=\alpha$ for each $i$ and $a$, finding
\begin{equation}\label{totale}
\arraycolsep=1pt\def\arraystretch{1.9}
\begin{array}{r@{}l}
\chi_{ij}^{(n)}&{}=\frac{1}{d_i}\frac{\alpha^n}{1+\sum_a\frac{\mu_a}{\mu_{0,ia}}} \, \left(\mathbf{X}^n\right)_{ij}~~,\\
X_{ij}&{}=\frac{m_i^\star}{\alpha\left(1+\sum_a\frac{\mu_a}{\mu_{0,ia}}\right)^2}\sum_{a} A_{ai}A_{aj}\frac{\mu_a^\star}{\mu_{0,ia}m_{0,ja}}\left(1+\sum_\ell \frac{m_\ell}{m_{0,\ell a}}\right)^2~~,
\end{array}
\end{equation}
where $\mathbf{X}$ is the matrix with elements $X_{ij}$ and $A_{ai}=1$ if miRNA $a$ targets ceRNA $i$ and zero otherwise. Because $\alpha<1$, one sees that the contribution coming from  chains of $n$ miRNA-mediated couplings becomes smaller and smaller (exponentially fast) as $n$ increases. Eq~(\ref{totale}) shows explicitly that ceRNAs $i$ and $j$ can crosstalk even when they have no miRNA regulator in common (in which case $X_{ij}=0$), provided there is a path of miRNA-mediated interactions connecting them (as suggested e.g. in \cite{figl,itza}; see also Fig. \ref{Fig5}). Hence, clearly, the topological structure of the miRNA-ceRNA network can strongly influence the emergent crosstalk scenario. The discussion presented here does virtually nothing to address the ensuing complexity. A deeper understanding of the interplay between topological and kinetic heterogeneities might shed light on the evolutionary drivers of miRNA targeting patterns and of the ceRNA mechanism.

\subsection{Noise processing}

\subsubsection{Noise buffering in small regulatory motifs}

Together with transcription factors (TFs), miRNAs form a highly interconnected network whose structure can be decomposed in small regulatory patterns, or circuits. Few of them, hereafter call {\it motifs}, are overrepresented and thus expected to perform regulatory functions. In particular, it has been proven that all these miRNA-mediated motifs play some role in stabilizing the expression of the miRNA-target against fluctions \cite{osella11, bosia12, riba14, osella14, grigolon16}. Amongst  others, a special role is performed by feedforward loops involving one miRNA, one TF and one target. Both the miRNA and the TF can play the role of the master regulator, while the target is dowregulated by the miRNA and activated or inhibited by the TF. 
The incoherent version of this motif, where the TF activates the expression of miRNA and target, can couple fine-tuning of the target together with an efficient noise control \cite{osella11, grigolon16}. Intuitively, this can be understood by noting that fluctuations that propagate from TF to target and miRNA are correlated, so that an increase or decrease in the amount of miRNA will coincide with a decrease in the amount of target.
The theoretical framework for the analysis of these effects is that of the ME, which in this case takes into account five different variables, one for each of the involved molecular species (mRNA and protein for the TF, mRNA and protein for the target, and the miRNA). The transcriptional activation of miRNA and target is modelled via a non-linear increasing Hill function of the number of TF, i.e. 
\begin{equation} 
\label{hill_activ}
b_m(f) = \frac{b_m f^c}{h_{m}^{c}+f^{c}} ~~~~~ , ~~~~~ \beta_{\mu}(f) = \frac{\beta_{\mu} f^c}{h_{\mu}^{c}+f^{c}} ~~ ,
 \end{equation}
where $b_m$ and $\beta_{\mu}$ are the transcription rates of target $m$ and miRNA $\mu$ respectively, $c$ is the Hill coefficient setting the steepness of the sigmoidal function and $h_{m}$ and $h_{\mu}$ are the dissociation constants, that specify the amount
of TF proteins $f$ at which the transcription rate is half of its maximal value ($b_m$ and $\beta_{\mu}$ respectively).
The miRNA interaction can be either modelled via a repressive Hill function of the number of miRNA molecules, i.e. as $b_f(\mu) = \frac{b_p h^{c}}{h^{c}+\mu^{c}}$, or via a titration-based mechanism. In the Hill function, $c$ is again the Hill coefficient and $h$ set the amount of miRNAs necessary to halve the maximum target translation rate $b_f$.
In the first case, one implicitly assumes that the miRNA action is catalytic (that is, the miRNA is never affected by the interaction with the target) and directs translational repression. In the second case, instead, one assumes that the miRNA action is stoichiometric, via binding and unbinding reactions (with rates $k_{m\mu}^{+}$ and $k_{m\mu}^{-}$ respectively). As long as miRNA and target mRNA are bound, the target cannot be translated. The miRNA might be affected by the interaction with the target (with recycling rate $\alpha$) and the target itself has an effective degradation rate that depends on the binding and unbinding rates and that is bigger than its intrinsic value $d_i$. In this case the miRNA actively promotes the degradation of the target. It is possible to show analytically and by numerical simulations that the maximal noise attenuation for the target is obtained for a moderate miRNA repression, independently of the way the miRNA interaction is modelled \cite{osella11}. This prediction, besides being in agreement with experimental observations of the impact of a wide
class of microRNAs on their target proteins, also suggests that an optimal noise reduction might be achieved even when the miRNA repression is diluted over multiple targets, provided these ceRNAs are not too noisy.

The analysis of data from the {\it Encyclopedia of DNA Elements} (ENCODE) \cite{encode} revealed that two other classes of miRNA-mediated circuits are enriched over the mixed network of miRNAs and TFs. One of them has a miRNAs that regulates two different genes that can eventually dimerize; the second has a miRNA that interacts with two TFs which in turn regulate the same gene. In both cases, the miRNA seems to have a role in stabilizing the relative concentration of their targets. The interesting fact is that a further enrichment appears when looking for those circuits in which there is a transcriptional connection between the two miRNA targets, i.e. one of them is a TF of the other. This TF, together with the miRNA, can in turn regulate multiple targets. This motif is again a feedforward loop where the miRNA plays the role of the master regulator and the TF and targets are ceRNAs. When modelling the motif with a titrative interaction for miRNA and target, in line with (\ref{eq:steadystateM}), and with an activatory Hill function from the TF to the target, it becomes clear that the topology of the circuit, together with the ceRNA interaction, enhances the coordination of the targets \cite{riba14}. This aspect is useful when TF and target have to maintain a fixed concentration ratio, which might be the case when they interact under a given stoichiometry.

\subsubsection{Transcriptional noise and the role of transcriptional correlations}

miRNA-mediated crosstalk can also provide a pathway to processing extrinsic noise, specifically cell-to-cell variability in transcription rates. Generalizing the lines that brought us to (\ref{cov}), one can say that if such a noise is sufficiently small, each component $\avg{x_k}$ of the steady state concentration vector $\avg{\mathbf{x}}= (\{\avg{m_i}\}_{i=1}^N,\{\avg{\mu_a}\}_{a=1}^M)$ can be written as
\begin{equation}\label{pob}
\avg{x_k} \simeq \overline{\avg{x_k}}+\sum_{s} \chi_{ks}(r_{s}-\overline{r}_{s})~~~~~,~~~~~\chi_{ks}=\frac{\partial \avg{x_k}}{\partial r_{s}}~~,
\end{equation}
where $\ovl{\avg{\mathbf{x}}}$ stands for the mean steady state vector (averaged over transcriptional noise), $r_s$ denotes the components of the vector $\mathbf{r}=(\{b_i\}_{i=1}^N,\{\beta_a\}_{a=1}^M)$ of transcription rates (including both those relative to ceRNAs and miRNAs), and the sum runs over all ceRNA and miRNA species. In turn, transcriptional noise induces fluctuations in the level of molecular species $k$ described by  \cite{figl}
\begin{equation} 
\label{sigma}
 \sigma^2_{k}\equiv\overline{(\avg{x_k}- \overline{\avg{x_k}})^2}= \sum_{s,s'} \chi_{ks}\,\chi_{ks'}\,\Sigma_{ss'}~~,
 \end{equation}
where $\boldsymbol{\Sigma}$ denotes the covariance matrix of transcription rates. If  $\boldsymbol{\Sigma}$ is diagonal, i.e. if transcription rates are mutually independent, the above expression reduces to
\begin{equation}
\sigma^2_{k}=\sum_{s} \chi_{ks}^2\, \Sigma_{ss}~~.
\end{equation}
This means that, in absence of transcriptional correlations, each molecular species in the network (both ceRNAs and miRNAs) contributes a positive quantity to the overall level of noise affecting species $k$. In such conditions, the latter clearly exceeds the intrinsic noise level $\Sigma_{kk}$. In particular, large competition-driven susceptibilities (both to perturbations affecting ceRNAs and to perturbations affecting miRNAs) may cause $\sigma_k^2$ to be much larger than $\Sigma_{kk}$, eventually leading to a loss of resolution in molecular levels that will necessarily limit crosstalk effectiveness.

\begin{figure}[t]
\begin{center}
\includegraphics[width=16cm]{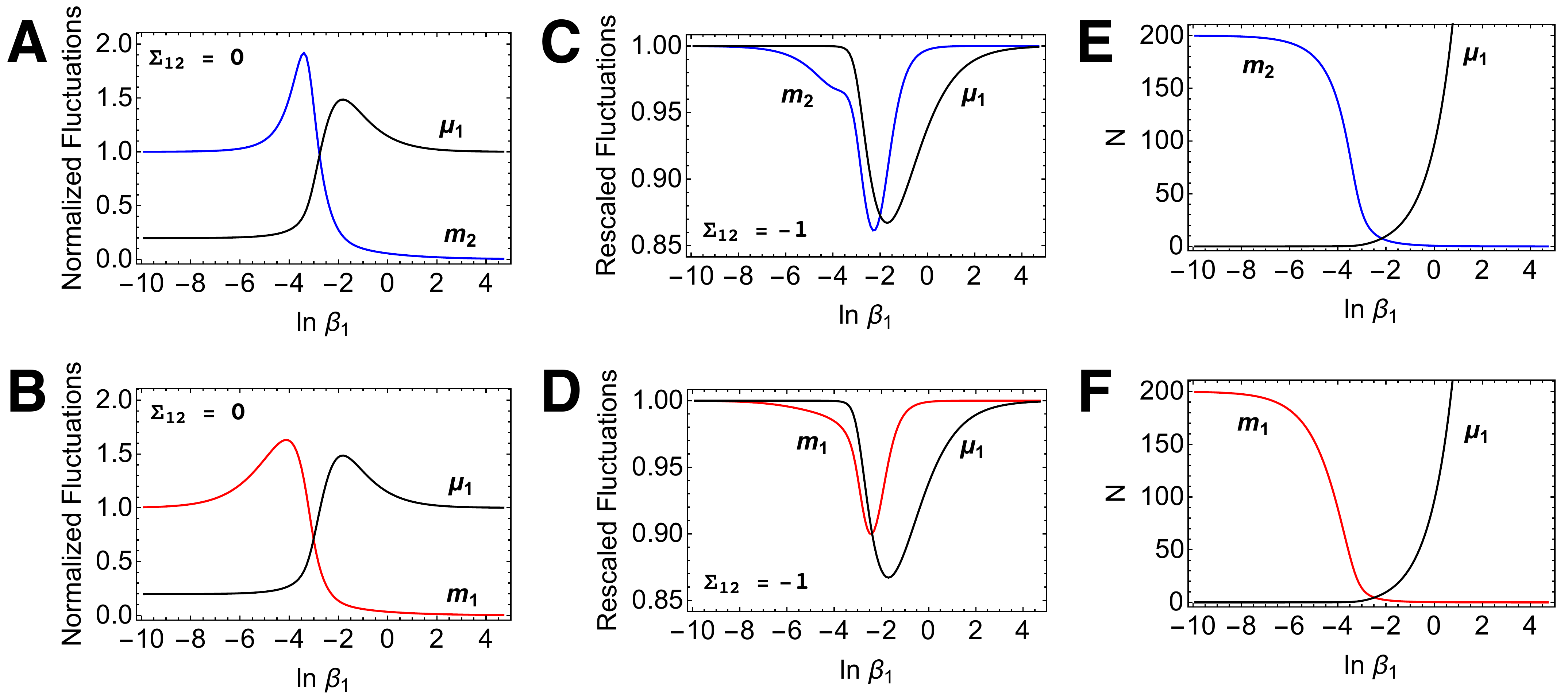}
\end{center}
\caption{{\bf (A, B)} Ratio between the magnitude of fluctuations for each molecular species for an interacting ($k_{11}^+,k_{21}^+>0$) and a non-interacting ($k_{11}^+,k_{21}^+=0$) system with 2 ceRNAs and one miRNA (``normalized fluctuations'') for uncorrelated ceRNA transcription rates ($\Sigma_{12}=0$) as a function of the miRNA transcription rate. {\bf (C, D)} Ratio between the normalized fluctuations obtained for maximally anti-correlated transcription rates ($\Sigma_{12}=-1$) and for the fully uncorrelated case ($\Sigma_{12}=0$) as a function of the miRNA transcription rate. {\bf (E, F)} Mean steady state molecular levels  as a function of the miRNA transcription rate. All results were obtained by averaging steady state solutions over transcriptional noise. Parameter values are reported in Table \ref{pars}.}
\label{Fig4}       % Give a unique label
\end{figure}

Interestingly, though, Eq (\ref{sigma}) suggests that the presence of transcriptional correlations (i.e. of off-diagonal terms in $\boldsymbol{\Sigma}$) can compensate for this effect \cite{figl}. For instance, negative correlations between ceRNA transcription rates tend to reduce the overall noise level affecting ceRNA $k$ with respect to the fully uncorrelated case (since both $\chi_{ks}$ and $\chi_{ks'}$ are non-negative if $k$, $s$ and $s'$ are ceRNAs). The same holds for positive correlations between the transcription rates of miRNAs and ceRNAs. In both cases, specific patterns of transcriptional correlations coupled with competition may confer a miRNA-ceRNA network the ability to buffer extrinsic noise. On the contrary, anti-correlated miRNA-ceRNA transcription rates or positively correlated ceRNA transcription rates tend to amplify extrinsic noise. These effects are displayed in Fig. \ref{Fig4}, where we show the fluctuation picture arising when all transcription rates are Gaussian distributed, with a fixed ratio between the average and the width.  
Uncorrelated ceRNA transcription rates lead to an enhancement of   fluctuations with respect to the case in which the miRNA is absent, while anti-correlated ceRNA transcription rates can attenuate this effect.

The noise-processing capacity of crosstalk patterns, and hence ultimately their effectiveness, is therefore strongly linked to the statistics of transcription rates. We shall see below that such correlations can indeed be exploited for the stabilization of the expression levels of protein complexes via the ceRNA mechanism.

\subsubsection{Emergence of bimodal gene expression}

As shown above, one of the main properties of molecular sequestration is the possibility to obtain threshold responses and ultrasensitivity in absence of molecular cooperativity (a property found also when one or more genes are regulated by miRNAs). We also recalled that the system Eq (\ref{uno}) possesses a unique, asymptotically stable steady state \cite{flon}. However, both theoretical and experimental studies have shown that miRNAs, in peculiar conditions of stoichiometry, induce bimodal distributions in the expression level of their targets \cite{bose12, bosi, sgro}. As reviewed in \cite{tsimring14} and shown in \cite{samoilov05}, some biological systems may present bimodality just as a consequence of stochasticity and despite being monostable at the deterministic level. The titrative interaction between miRNA and targets places targets, and ceRNAs in general, into this class of systems. Indeed, when the target expression level is around the threshold established by the amounts of miRNA, if the interaction is sufficiently strong, a small fluctuation in the amount of miRNA or target molecules makes the system jump from the repressed to the unrepressed regime and viceversa. The direct outcome is a bimodal distribution of the targets around the threshold, whose modes are related to the repressed and unrepressed regimes. %Such a bimodal distribution, which is so far solely due to the intrinsic stochasticity of the involved chemicals, can propagates to the ceRNAs in proximity of the same threshold if also their interactions with the miRNA is strong enough. 

The constraint of strong miRNA-target interaction can however be relaxed by introducing some extrinsic noise in the system. This scenario has been exhaustively addressed, both analytically and numerically, in \cite{delgiudice18}. Let us focus on a simple system with two ceRNAs and one miRNA. The system is described by the probability distribution $P(\mu,m_1,m_2,t|\mathbf{K})$ of observing $\mu$ molecules of miRNAs and $m_1,m_2$ molecules of mRNAs of target $1$ and $2$ at time $t$, for a given set of parameters $\mathbf{K} = \{b_1,b_2,\beta,d_1,d_2,\delta,k_{1\mu}^+,k_{2\mu}^+\}$. Such a probability distribution evolves according to the ME (\ref{eq:masterCompleta}) with $N=2$ and $M=1$. Fluctuations in $\mathbf{K}$ should be taken into account in order to obtain the full distribution at the steady state $P(\mu,m_1,m_2)$. %In order to do that, we need to recall the law of total probability, $P(\mu,m_1,m_2)=\int P(k) P(\mu,m_1,m_2|K) dK$. 
For sakes of simplicity, now assume that $\beta$ is the only fluctuating rate, drawn from a Gaussian distribution centered around $\avg{\beta}$, with variance $\sigma_{\beta}^{2}$ and defined for $\beta>0$. We can obtain the steady-state probability distribution $P(\mu,m_1,m_2|\beta)$ conditional on a specific $\beta$ by applying e.g. the LNA or the Gaussian approximation to the ME. Once this is done, the joint distribution $P(\mu,m_1,m_2)$ is found by performing a weighted average over all possible values of $\beta$, i.e. by applying the law of total probability: $P(\mu,m_1,m_2)=\int P(\beta) P(\mu,m_1,m_2|\beta) d\beta$. 

The presence of extrinsic noise in terms of fluctuating parameters is such that the miRNA transcription rate $\beta$ is not the same for every cell as for the pure intrinsic noise case (indeed, we are extracting $\beta$ from a Gaussian distribution). This implies that picking values of $\beta$  above or below the threshold has the consequence of placing the system in the repressed or unrepressed regime respectively. Again, the outcome is a bimodal distribution, which is this time at the population level. Then, the larger the variance $\sigma_{\beta}^2$ (i.e. the extrinsic noise), the broader the ranges of expressed target explored by the left-tails of the Gaussian distribution that will superimpose in the unrepressed mode. The right-tail instead will accumulate cells in the repressed mode. This makes the threshold/noise coupling an efficient tool to filter the variability introduced by extrinsic noise.

\subsubsection{Impact on protein expression}
\label{prots} 

\begin{figure}[t]
\begin{center}
\includegraphics[width=16cm]{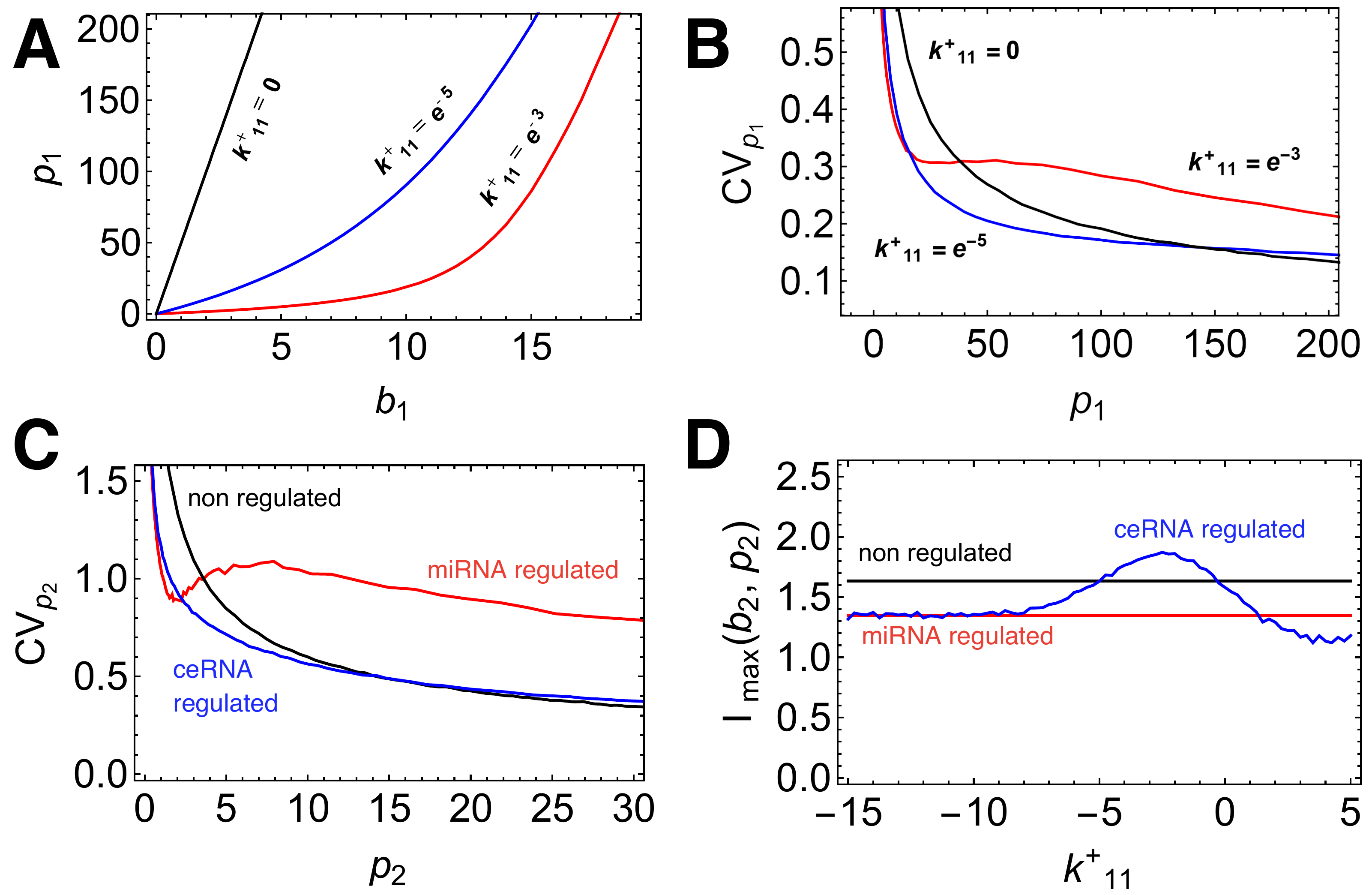}
\end{center}
\caption{{\bf (A)} Mean level of a protein ($p_1$) interacting with a miRNA versus the transcription rate of its mRNA ($b_1$) for different values of the miRNA-mRNA interaction strength. No competitor is present. Expression of $p_1$ gets a stronger threshold-linear behaviour as the miRNA-mRNA interaction strength increases. {\bf (B)} Relative fluctuations (CV)  of $p_1$ versus $b_1$, again in absence of competition. {\bf (C)} CV of a target protein ($p_2$) as a function of the mean protein level for the case in which the protein is not interacting with a miRNA (black line, $k_{11}^+=k_{21}^-=0$), is miRNA-regulated but has no competitor (red line, $k_{11}^+=0$, $k_{21}^+=1$) and is miRNA-regulated and has a competitor (blue line, $k_{11}^+=e^{-2}$, $k_{21}^+=1$). {\bf (D)} Maximal mutual information between $p_2$ and its transcription rate $b_2$ for the three regulatory modes presented in (C), plotted as a function of the miRNA-competitor interaction strength. The ceRNA-effect provides the most efficient fine-tuning pathway for intermediate strengths. We used $g_1=0.5$/min and $q_1=0.1$/min for panels (A) and (B); $g_1 = g_2= 0.5$/min, $q_1 = q_2=0.1$/min
for panels (C) and (D). Remaining parameter values are reported in Table \ref{pars}.}
\label{Fig6}       % Give a unique label
\end{figure}

The ability of generic regulatory systems to process noise is most crucial for the fine tuning of protein levels \cite{lope}. Interestingly, the control exterted by miRNAs on a single target  has been found to be capable of buffering its expression noise \cite{goya}, especially for sufficiently low expression levels \cite{schm}. Given this scenario, one can ask whether the presence of a competitor would improve noise processing, especially at high expression, with the rationale that fluctuations affecting the target mRNA will be smaller (at fixed average) if a competitor titrates regulatory miRNAs away from it. This idea has been tested in simulations after modifying the basic model, Eq. (\ref{due}), to account for protein production \cite{cern}. This is done by simply including the extra equation 
\begin{equation}
\frac{d p_i }{dt }= g_i m_i - q_i p_i~~,
\end{equation}
which, for each mRNA species $i$, describes the time evolution of the level $p_i$ of proteins of type $i$ due to synthesis  (occurring at rate $g_i$ per substrate molecule) and degradation (occurring at rate $q_i$ per protein). Fluctuations affecting $p_i$ depend on the strength of the interaction between the miRNA and the target's competitor. A weak coupling is insufficient to draw miRNAs away from the target, leading (expectedly) to the same qualitative picture found in absence of the competitor. Likewise, very strong miRNA-competitor coupling leaves the target free from miRNAs, in which case its noise level is comparable to that attained in absence of miRNAs. However, for an intermediate value of the miRNA-competitor binding rate, titration by the competitor appears to be optimally tuned to reduce target fluctuations even at high expression levels (see Fig. \ref{Fig6}). 
In this regime, the competitor is maximally derepressed.  Remarkably, the overall behaviour of relative fluctuations is  close to the Poissonian scenario obtained for an unregulated protein, implying that target derepression plays the main role in reducing fluctuations. Moreover, when crosstalk is most efficient, noise at low expression levels is still efficiently buffered with respect to the case in which miRNAs are absent. A more refined analysis shows that miRNA recycling generically provides enhanced fine tuning by increasing the effective miRNA level.

The fact that, in the human PPI network, the functional products of mRNAs targeted by the same miRNAs are more strongly connected than would be expected by chance strongly suggests that miRNA-mediated regulation, and by extension the ceRNA mechanisms, might play a role in the regulation of protein complex levels \cite{lian,yuan,sass}. In particular, protein forming the subunits of larger complexes tend to be regulated by miRNA clusters, i.e. by groups of miRNA species that are co-expressed \cite{hsu}. When competing RNAs are the substrate for the synthesis of interacting proteins, the onset of the ceRNA mechanism modifies the correlation pattern of the two sub-units, specifically changing the sign of correlations from negative (corresponding to sub-units that are not co-regulated) to positive (reflective the positive correlation that is established between ceRNAs in crosstalk conditions). Such a modification has been observed experimentally \cite{du,nada,kwon}, suggesting that it might provide a biological (albeit non-universal) signature of the ceRNA effect in action.

\subsubsection{Limits to crosstalk effectiveness}

From the previous discussion it is clear that the effectiveness of the ceRNA mechanism is dictated in large part by the relative levels of the molecular species involved and is ultimately limited by noise. An important question in this respect is whether one can characterize the optimal performance that miRNA-mediated regulation can achieve in controlling gene expression. In general, the optimal properties achievable by a regulatory circuit describe fundamental physical limits to its performance,  which cannot be overcome independently of kinetic details, and point to the individual processes constituting, in some sense, the bottlenecks for regulatory effectiveness. It is clear that this requires, on one hand, a quantitative definition of `regulatory effectiveness' and, on the other, a benchmark. To fix ideas, one can focus on the system formed by a single miRNA connecting two competing RNAs. Following \cite{tkac}, a natural definition for the effectiveness of ceRNA crosstalk is represented by the degree to which one can control the level of one of the ceRNAs, say ceRNA $i$, by modulating the level of its competitor (ceRNA $j$). In a stochastic setting, the miRNA-mediated interaction linking $i$ and $j$ can be seen as a ``communication channel'' that probabilistically translates the transcription rate of $j$ into a value of $m_i$. This channel is fully described by the conditional probability density $p(m_i|b_j)$, returning a random value of $m_i$ (the contributing noise coming from all involved processes) upon presenting input $b_j$. In turn, miRNA-mediated regulation consists in processing, via $p(m_i|b_j)$ a distribution of values for $b_j$ (denoted by $p(b_j)$) into a distribution of values of $m_i$. For any given $p(m_i|b_j)$ and $p(b_j)$, the strength of the mutual dependence between these variables is quantified by the mutual information \cite{prob}
\begin{equation}
I(b_j,m_i)=\int_{b_j^\minn}^{b_j^{\maxx}} db_j ~ p(b_j)\int_{{m_i^\minn}}^{{m_i^\maxx}} dm_i ~ p(m_i|b_j) \log_2 \frac{p(m_i|b_j)}{p(m_i)}~~,
\end{equation}
with $p(m_i)=\int_{b_j^\minn}^{b_j^{\maxx}} db_j p(m_i|b_j)p(b_j)$ the output distribution of $m_i$. Assuming that the channel is fixed, i.e. that $p(m_i|b_j)$  is given, the optimal regulatory effectiveness is obtained when the input distribution $p(b_j)$ is such that $I$ is maximized:
\begin{equation}
\max_{p(b_j)} ~ I (b_j, m_i) \equiv I_{\max} ~~.
\end{equation}
$I_{\max}$ is called the {\it capacity} in information-theoretic terms and ultimately measures how much information (in bits) can be conveyed at most from input ($b_j$) to output ($m_i$) by a given input-output relationship $p(m_i|b_j)$. In loose but intuitive terms, $I_\maxx$ describes the number $\mathcal{N}$ of different values $m_i$ that can be  distinguished in a reliable way given the noise, which is roughly given by $\mathcal{N}\sim 2^{I_\maxx}$. If $I_\maxx\simeq 0$ (note that $I\geq 0$ by definition), the noise only allows to distinguish at most one level of $m_i$; for $I_\maxx\simeq 1$ two levels (high/low) can be separated; and so on.  

The effectiveness of miRNA-mediated crosstalk has been characterized within the above setup starting from numerical simulations of the stochastic dynamics and using direct transcriptional regulation of $m_i$ (i.e., the capacity of the corresponding miRNA-independent regulatory channel) as the benchmark against which miRNA-mediated information flow was evaluated. In particular, the dependence of $I_\maxx$ on kinetic parameters was analyzed to identify optimal parameter regions and limiting processes. The emergent scenario can be summarized as follows \cite{prob}:
\begin{enumerate}
\item As might have been expected, the capacity of miRNA-mediated regulation is optimal in a specific range of values for the target's repression strength. Intuitively, a tight control of $m_i$ based on $b_j$ requires ceRNA $i$ to be sensitive to changes in miRNA levels. Too weak (resp. too strong) repression causes ceRNA $i$ to become fully unrepressed (resp. fully repressed), so that the optimal range lies between these extremes. Quite remarkably, though, optimal ceRNA crosstalk can be more effective than direct transcriptional control. 
\item In presence of significantly different catalytic degradation rates (faster for $m_i$, slower for $m_j$) ceRNA crosstalk outperforms direct transcriptional regulation. Intuitively, the above situation makes transcriptional control especially inefficient since $m_i$ is going to be strongly repressed by miRNAs. miRNA-mediated control, instead, benefits  from the fact that $m_j$ can de-repress ceRNA $i$ by lifting miRNAs away from it.
\item When miRNA populations are sufficiently large and miRNA-ceRNA couplings are weak, miRNA-mediated regulation is as effective as a direct transcriptional control. This is intuitively due to the fact that, in this limit, the relative noise affecting miRNA levels becomes negligible. This removes the additional source of noise affecting the post-transcriptional channel compared to the transcriptional one, effectively making the two regulatory modes comparable.
\end{enumerate}
The outlook is that, besides generically contributing to noise buffering, ceRNA crosstalk can control of gene expression to a degree that is tightly connected to the ability of the competitor ($m_j$) to de-repress the target ($m_i$). When the controller's kinetics does not suffice to titrate miRNAs away from ceRNA $i$, miRNA-mediated regulation is ineffective. Otherwise, it provides a high (and, possibly, the highest achievable) degree of control over expression levels, especially when kinetic parameters are sufficiently heterogeneous. 

\subsection{ceRNA crosstalk away from stationarity}
\label{sec:3}
% Always give a unique label
% and use \ref{<label>} for cross-references
% and \cite{<label>} for bibliographic references
% use \sectionmark{}
% to alter or adjust the section heading in the running head
\subsubsection{Equilibration times}

\begin{figure}[t]
\begin{center}
\includegraphics[width=16cm]{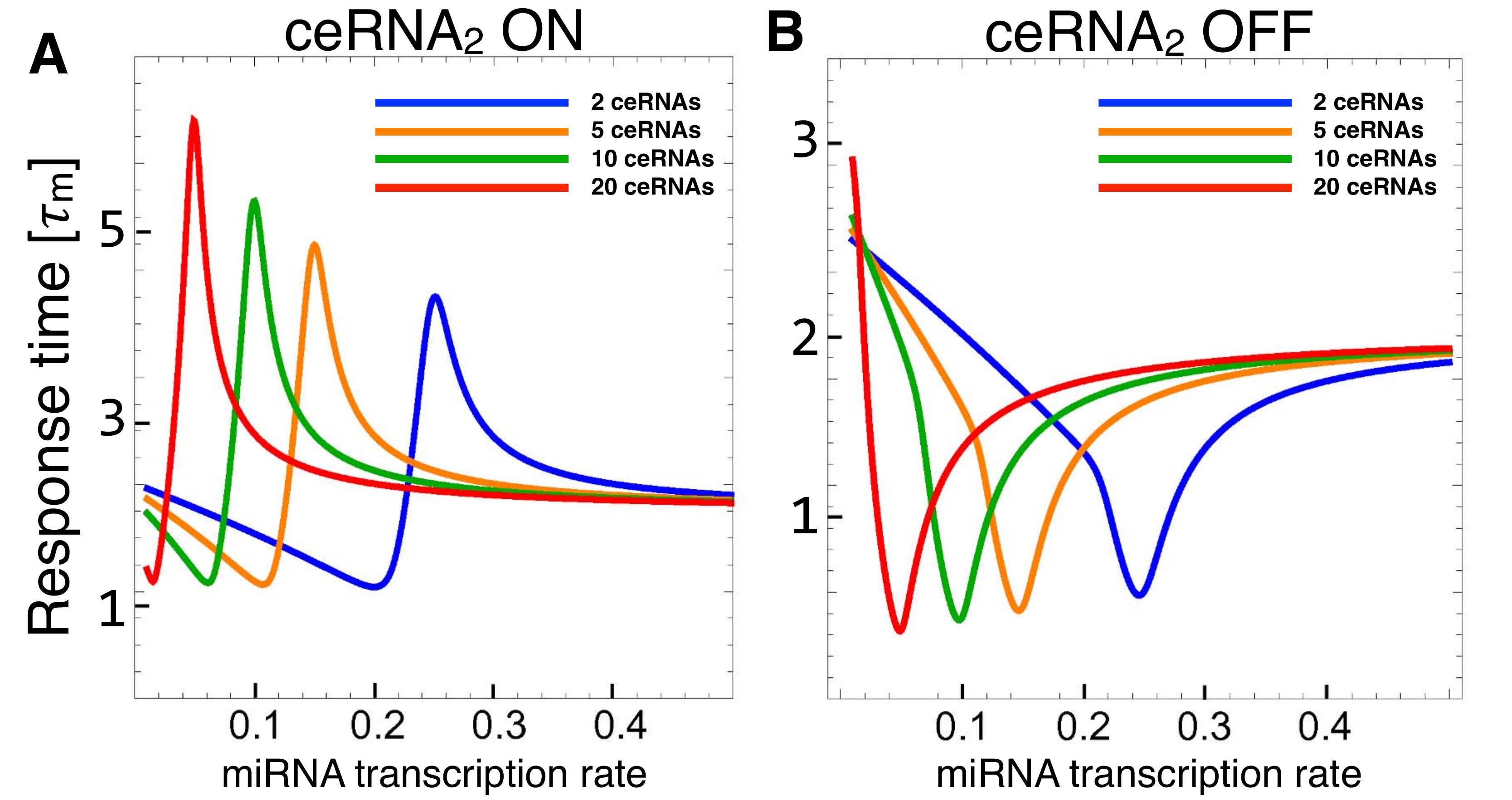}
\end{center}
\caption{We consider a simple ceRNA network for an increasing number of targets ranging from 2 to 20 and a single microRNA. {\bf (A)} Equilibration time of ceRNA$_1$ when ceRNA$_2$ is induced, as a function of the miRNA transcription rate. Around threshold, we observe a critical slowing down in the response time. {\bf (B)} Same as (A), but now the response time is measured after a knock-down of one of the competitors. In this case, we observe a speed up of the response time at threshold.}
\label{Fig7}       
\end{figure}

The titrative miRNA-target interaction entails both susceptibility and statistical correlation between the competing chemical species. We have seen before how all these effects become maximal at quasi equimolar ratio. One can however also study how fast the system responds to an external perturbation. To fix ideas, we will focus as usual on the case of a single miRNA targeting 2 ceRNAs. In particular, we want to quantify the time needed for a particular ceRNA (here ceRNA$_1$) to reach the new stationary state after
\begin{itemize}
\item A sudden increase of the transcriptional activity of ceRNA$_2$ at time $t=0$, i.e.
  \begin{displaymath}
    b_2(t=0^-)=0 \quad\mathrm{and}\quad b_2(t=0^+)=b^*
  \end{displaymath}
\item A sudden decrease of the transcriptional activity of ceRNA$_2$  at time $t=0$, i.e.
  \begin{displaymath}
    b_2(t=0^-)=b^* \quad\mathrm{and}\quad b_2(t=0^+)=0
  \end{displaymath}
\end{itemize}
We define the response time as the time needed for ceRNA$_1$ to reach half the way between the initial (before perturbation) and final (after perturbation) steady state levels. In particular one can evaluate the response times $T_\mathrm{ON}$ and $T_\mathrm{OFF}$ for both the switch-on and switch-off scenarios (i.e. for ceRNA$_2$ $\mathrm{OFF}\rightarrow\mathrm{ON}$ and $\mathrm{ON}\rightarrow\mathrm{OFF}$ respectively) by numerically integrating Eq.~(\ref{uno}) to estimate $T_\mathrm{ON/OFF}$ as the times at which the following relations hold: 
\begin{eqnarray}
m_1(T_\mathrm{ON}) &=& m_1(0)+\frac12 \left(\lim_{t\rightarrow\infty}
m_1(t) - m_1(0)\right) \quad m_1(0) = m_1^\mathrm{ss}~,~m_2(0) = 0 \\
m_1(T_\mathrm{OFF}) &=& m_1(0)-\frac12 \left(m_1(0) -
\lim_{t\rightarrow\infty} m_1(t)\right) \quad m_1(0) = m_1^\mathrm{ss}~,~m_2(0) = m_2^\mathrm{ss}\,
\end{eqnarray}
In this framework we can easily study the dependence of the response times $T_\mathrm{ON/OFF}$ on the basal miRNA concentration (i.e. on $\beta_1$ in this case). Results (see Fig.~\ref{Fig7}) show a non-monotonous dependence of $T_\mathrm{ON/OFF}$ on the trascriptional activity of the miRNA. In particular, $T_\mathrm{ON}$ (resp. $T_\mathrm{OFF}$) displays a maximum (resp. a minimum) in correspondence with the threshold between the repressed and unrepressed phase. %Interestingly, the extreme is reached in the phase where the distance between the pre-perturbation steady state.

A natural question is how the presence of more ceRNAs changes the scenario we just described for the simple one miRNA two ceRNAs network. The same {\em in silico} experiment can be generalized to an arbitrary number of ceRNAs where all but one (say ceRNA$_2$) is either knock-out or induced. Perhaps unsurprisingly (see Fig.~\ref{Fig7}), one again sees a dilution effect: upon increasing the number of ceRNAs from 2 to 20 the relevance of the effect --measured in terms of the distance between the initial and final state of the system-- becomes quantitatively less relevant.

\subsubsection{Out of equilibrium dynamics} 

%Up to this point we have considered a titrative stoichiometric miRNA-ceRNA interaction. We have seen in particular that, for purely catalytic miRNA-ceRNA complex processing, no crosstalk  can occur at stationarity. Is it still possible to observe competition far from equilibrium? To analyze this possibility, we considered the time evolution of Eq.~(\ref{uno}) analyzing snapshots of the system at different times. If the time interval considered is smaller than the equilibration time of the complexes, we observe threshold behavior qualitatively similar to the one observed at stationarity in Fig.~(\ref{Fig2}). However, the sharpness of the threshold behavior becomes less and less pronounced as time increases, until it disappears in the $t\rightarrow\infty$. In other terms, although no competition is mathematically possible at stationarity in the catalytic regime ($\alpha = 0$), we can observe the onset of a dynamical threshold whose value at a given time $t$ tends monotonously to the stationary one for $\alpha \neq 0$ and to infinity in case of $\alpha = 0$. In the latter case no cross-talk is observed at equilibrium. The ceRNA effect is  therefore robust also in case of catalytic miRNA-target interaction, provided the system is far from equilibrium.

The out-of-equilibrium dynamics of the miRNA-ceRNA system has been studied in \cite{dyna}. The emergent crosstalk scenario is substantially richer than the stationary one. For simplicity, we shall limit ourselves to describing results obtained for a system with $N$ ceRNAs interacting with and a single miRNA species. From a physical viewpoint, the quantities
\begin{equation}
\arraycolsep=1pt\def\arraystretch{1.8}
\begin{array}{r@{}l}
&{}\tau_0=\delta^{-1} \quad, \quad\tau_{1,i}=d_i^{-1}\quad ,\quad \tau_{2,i}=(\sigma_i+\kappa_i+k_i^-)^{-1} \\
&{}\tau_{3,i}=(\sigma_i+\kappa_i)^{-1} \quad ,\quad \tau_{4,i}=\sigma_i^{-1}~~, \quad \tau_{5,i}=\kappa_i^{-1}
\end{array}
\end{equation}
represent the relevant characteristic intrinsic time scales of this system. Based on Eqs (\ref{uno}), they represent, respectively, the mean lifetime of miRNA species $a$ ($\tau_0$)) and of ceRNA species $i$ ($\tau_{1,i}$), the mean lifetime of the complex formed by ceRNA $i$ ($\tau_{2,i}$), and the mean time required for complex degradation ($\tau_{3,i}$), stoichiometric complex degradation ($\tau_{4,i}$) and catalytic complex degradation ($\tau_{5,i}$). The features characterizing dynamical crosstalk can change depending on how these time scales are related. To get some insight, one can focus on how the system relaxes back to the steady state following a small perturbation away from it. Upon linearizing the system (\ref{uno}), one can derive equations for the deviations of each molecular species from the steady state, i.e. for the quantities
\begin{equation}
\arraycolsep=1pt\def\arraystretch{1.5}
\begin{array}{r@{}l}
x_i(t) &{}\equiv m_i(t)-\avg{m_i} \\
y(t)&{}\equiv \mu(t)-\avg{\mu}\\
z_i(t)&{}\equiv c_i(t)-\avg{c_i}
\end{array}~~.
\end{equation}
(We have suppressed the miRNA index for sakes of simplicity.) Introducing (small) time-dependent additive perturbations of the transcription rates of the form $\btil_i(t)$ and  $\betil(t)$, the above variables can be seen to evolve in time according to
\begin{equation}\label{dyx}
\arraycolsep=1pt\def\arraystretch{1.8}
\begin{array}{r@{}l}
\frac{d}{dt}x_i&{}=-d_ix_i + \btil_i-k_i^+(\mu\,x_i+m_i\,y)+k_i^-z_i\\
\frac{d}{dt}y&{}=- \delta y + \betil-\sum_i k_i^+(\mu\,x_i+m_i\,y)+\sum_i(k_i^-+\kappa_i)z_i  \\
\frac{d}{dt}z_i&{}=-(\sigma_i+k_i^-+\kappa_i)z_i + k_i^+(\mu\,x_i+m_i\,y)
\end{array}~~,
\end{equation}
This system can be analyzed in the frequency domain ($\omega$) by Fourier-transforming (\ref{dyx}). This allows to define the {\it dynamical susceptibility}
\begin{equation}\label{susk1}
\widehat{\chi_{ij}}(\omega)=\frac{\partial \widehat{x_i}}{\partial \widehat{\btil_j}}~~,%=
\end{equation}
where $\widehat{f}$ is the Fourier transform of $f$. The general study of this quantity is possible while not straightforward \cite{detw}. However, $\widehat{\chi_{ij}}(\omega)$ can be estimated in a relatively simple way in few instructive limiting cases in which timescales are sufficiently separated. For instance, when $\tau_{3,j}\ll 1/k_j^-$ and $\tau_{1,j}<\tau_{5,j}$, complexes formed by ceRNA $j$ will typically keep miRNAs blocked for times longer than the intrinsic ceRNA degradation timescale. This may allow for ceRNA $i$ to get de-repressed and hence for the establishment of crosstalk, independently of whether stoichiometric processing takes place. Indeed one finds that, when $\kappa_j \ll \omega\ll d_j$ (i.e. for timescales intermediate between $\tau_{1,j}$ and $\tau_{5,j}$), 
\begin{equation}\label{qq}
\widehat{\chi_{ij}}(\omega)\simeq 
%\lim_{\substack{\sigma_j \to\kappa_j\\ \kappa_j \to 0}}
\frac{\sigma_j+\kappa_j}{\sigma_j}~\chi_{ij}~~
\end{equation}
where $\chi_{ij}=\frac{\partial\avg{m_i}}{\partial b_j}$ stands for the steady-state susceptibility \cite{dyna}. Remarkably, the quantity on the left-hand-side of Eq (\ref{qq}) can be shown to remain finite for $\sigma_j\to 0$, providing quantitative support to the observation that ceRNA crosstalk can be active dynamically even in purely catalytic systems (where no crosstalk occurs at stationarity and $\chi_{ij}$ vanishes). In other words, then, in this limiting case and in an intermediate frequency window, the dynamical susceptibility is comparable to the steady-state value and occurs even for $\sigma_j=0$. Away from this window, instead, crosstalk in this limit is weaker than it is at stationarity.

A more careful analysis shows that, in certain regimes, the dynamical response can even exceed the stationary one. This happens, for instance, when complex dissociation is much faster than other processing pathways and ceRNAs are fully repressed, implying that dynamical crosstalk can occur even between pairs of ceRNAs that could not interact at steady state \cite{dyna}. In this sense, the ceRNA mechanism out of equilibrium is substantially more complex and richer than its stationary counterpart. In addition, the possibility to modulate the time scales of different interactions allows to construct systems in which static and dynamic responses are tuned so as to ensure the correct transient activation of a specific gene and the long-term stabilization of expression levels. An example of such a coordination, based on findings related to skeletal muscle cell differentiation \cite{legn}, has bee studied in \cite{fior}.

\section{Outlook}

Mathematical models developed to elucidate the emergent features of ceRNA crosstalk have so far mainly relied on computational schemes for stochastic simulations (Gillespie algorithm) and on analytical approximations of the master equation associated to the system of interacting molecules (LNA, Gaussian, Langevin). On the other hand, a full understanding of competition-driven coupling requires, as we have seen, disentangling it from concurrent effects. Indeed, the identification of crosstalk from transcriptional data is in our view especially hard since statistical correlations between RNAs sharing a common miRNA regulator can arise just due to the fact that they both respond to fluctuating miRNA levels. Once the relationship between competition- and fluctuations-related features is clarified, ceRNA crosstalk patterns display strong intrinsic specificities like 
\begin{enumerate}
\item selectivity, 
\item asymmetry, 
\item plasticity (i.e. sensitivity to kinetic parameters), 
\item sensitivity to the degree of parameter heterogeneity, and 
\item the possibility to aggregate a large number of weak interactions to significantly impact molecular levels. 
\end{enumerate}
These features in turn allow for the establishment of complex noise-processing properties. Note that, unsurprisingly, some of these features characterize other competition scenarios in regulatory system (e.g. competition to bind transcription factors, $\sigma$-factors, ribosomes, etc. \cite{maur,brew,tull}).

We have reviewed these aspects together with the methods that can be employed to quantify them. Several important points might however deserve equal consideration. In first place, miRNAs can also crosstalk through ceRNAs, generating a very similar phenomenology whose impact has been, to our knowledge, far less clarified \cite{loin}. Secondly, the modeling framework we discussed ignores some kinetic steps assuming essentially that they are non rate-limiting. Still, it is known that in some cases binding to Argonaute (Ago), the catalytic component of the RNA-induced silencing complex, represents a kinetic bottleneck \cite{koll}. Likewise, crosstalk can be affected by the competition to bind Ago \cite{loin}. Third,   a rich trafficking of miRNAs and their targets is known to occur between the cell nucleus and the cytoplasm, leading to remarkable localization effects whose biological significance is largely unexplored \cite{nils}. Well-mixed models like those discussed here are clearly unable to deal with such effects; spatial generalizations are mandatory \cite{levlev,levi}. Finally, the phenomenology derived from small modules can integrate in highly non-trivial ways at the scale of the transcriptome, where topology provides additional degrees of freedom to modulate crosstalk patterns. While, as shown here, some (basic) things about the role of network structure can be understood with simple calculations, a more thorough data-based analysis of these aspects would be greatly welcome.

\begin{acknowledgement}
Work supported by the European Union's Horizon 2020 research and innovation programme MSCA-RISE-2016 under grant agreement No 734439 INFERNET. We are indebted with Matteo Figliuzzi, Enzo Marinari, Matteo Marsili and Riccardo Zecchina for our fruitful and enjoyable collaboration.
\end{acknowledgement}

%\footnotesize{
%\tableofcontents
%}

\begin{thebibliography}{999}

\bibitem{bartel}Bartel DP (2009) MicroRNAs: target recognition and regulatory functions. Cell 136:215-33.

\bibitem{flynt}Flynt AS, Lai EC (2008) Biological principles of microRNA-mediated regulation: shared themes amid diversity. Nature Reviews Genetics 9:831.

\bibitem{cech}Cech TR, Steitz JA (2014) The noncoding RNA revolution--trashing old rules to forge new ones. Cell 157:77-94.

\bibitem{gurt}Gurtan AM, Sharp PA (2013) The role of miRNAs in regulating gene expression networks. Journal of Molecular Biology 425:3582-600.

\bibitem{metaz} Bartel DP (2018) Metazoan microRNAs. Cell 173:20-51.

\bibitem{risc}Gregory RI, Chendrimada TP, Cooch N, Shiekhattar R (2005) Human RISC couples microRNA biogenesis and posttranscriptional gene silencing. Cell 123:631-40.

\bibitem{chan}Chandradoss SD, Schirle NT, Szczepaniak M, MacRae IJ, Joo C (2015) A dynamic search process underlies microRNA targeting. Cell 162:96-107.

\bibitem{why}Klein M, Chandradoss SD, Depken M, Joo C (2017) Why Argonaute is needed to make microRNA target search fast and reliable. In Seminars in Cell \& Developmental Biology  (Vol. 65, pp. 20-28). Academic Press.

\bibitem{chek}Chekulaeva M, Filipowicz W (2009) Mechanisms of miRNA-mediated post-transcriptional regulation in animal cells. Current Opinion in Cell Biology 21:452-60.

\bibitem{jona}Jonas S, Izaurralde E. Towards a molecular understanding of microRNA-mediated gene silencing (2015) Nature Reviews Genetics 16:421.

\bibitem{djur}Djuranovic S, Nahvi A, Green R (2012) miRNA-mediated gene silencing by translational repression followed by mRNA deadenylation and decay. Science 336:237-40.

\bibitem{bart}Bartel DP (2004) MicroRNAs: genomics, biogenesis, mechanism, and function. Cell 116:281-97.

\bibitem{liang}Liang Y, Ridzon D, Wong L, Chen C (2007) Characterization of microRNA expression profiles in normal human tissues. BMC genomics 8:166.

\bibitem{fran}Franks A, Airoldi E, Slavov N (2017) Post-transcriptional regulation across human tissues. PLoS Computational Biology 13:e1005535.

\bibitem{eber}Ebert MS, Sharp PA (2012) Roles for microRNAs in conferring robustness to biological processes. Cell 149:515-24.

\bibitem{bere}Berezikov E (2011) Evolution of microRNA diversity and regulation in animals. Nature Reviews Genetics 12:846.

\bibitem{josh}Joshi A, Beck Y, Michoel T (2012) Post-transcriptional regulatory networks play a key role in noise reduction that is conserved from micro-organisms to mammals. The FEBS Journal 279:3501-12.

\bibitem{frie}Friedman RC, Farh KK, Burge CB, Bartel DP (2009) Most mammalian mRNAs are conserved targets of microRNAs. Genome Research 19:92-105.

\bibitem{baek}Baek D, Vill\'en J, Shin C, Camargo FD, Gygi SP, Bartel DP (2008) The impact of microRNAs on protein output. Nature 455:64.

\bibitem{shim}Shimoni Y, Friedlander G, Hetzroni G, Niv G, Altuvia S, Biham O, Margalit H (2007) Regulation of gene expression by small non-coding RNAs: a quantitative view. Molecular Systems Biology 3:138.

\bibitem{tsang}Tsang J, Zhu J, van Oudenaarden A (2007) MicroRNA-mediated feedback and feedforward loops are recurrent network motifs in mammals. Molecular Cell 26:753-67.

\bibitem{reda}Re A, Cor\'a D, Taverna D, Caselle M (2009) Genome-wide survey of microRNA-transcription factor feed-forward regulatory circuits in human. Molecular BioSystems 5:854-67.

\bibitem{sici}Siciliano V, Garzilli I, Fracassi C, Criscuolo S, Ventre S, Di Bernardo D (2013) MiRNAs confer phenotypic robustness to gene networks by suppressing biological noise. Nature Communications 30:2364.

\bibitem{wangs}Wang S, Raghavachari S. Quantifying negative feedback regulation by micro-RNAs (2011) Physical Biology 8:055002.

\bibitem{das}Das J, Chakraborty S, Podder S, Ghosh TC (2013) Complex-forming proteins escape the robust regulations of miRNA in human. FEBS Letters 587:2284-7.

\bibitem{ober}Obermayer B, Levine E (2014) Exploring the miRNA regulatory network using evolutionary correlations. PLoS Computational Biology 10:e1003860.

\bibitem{schm}Schmiedel JM, Klemm SL, Zheng Y, Sahay A, Bl\"uthgen N, Marks DS, van Oudenaarden A (2015) MicroRNA control of protein expression noise. Science 348:128-32.

\bibitem{guil}Guil S, Esteller M (2015) RNA-RNA interactions in gene regulation: the coding and noncoding players. Trends in Biochemical Sciences 40:248-56.

\bibitem{hans}Hansen TB, Jensen TI, Clausen BH, Bramsen JB, Finsen B, Damgaard CK, Kjems J (2013) Natural RNA circles function as efficient microRNA sponges. Nature 495:384.

\bibitem{eber2}Ebert MS, Neilson JR, Sharp PA. MicroRNA sponges: competitive inhibitors of small RNAs in mammalian cells (2007) Nature Methods 4:721.

\bibitem{suma}Sumazin P, Yang X, Chiu HS, Chung WJ, Iyer A, Llobet-Navas D, Rajbhandari P, Bansal M, Guarnieri P, Silva J, Califano A (2011) An extensive microRNA-mediated network of RNA-RNA interactions regulates established oncogenic pathways in glioblastoma. Cell 147:370-81.

\bibitem{helw}Helwak A, Kudla G, Dudnakova T, Tollervey D (2013)  Mapping the human miRNA interactome by CLASH reveals frequent noncanonical binding. Cell 153:654-65.

\bibitem{kimd}Kim D, Sung YM, Park J, Kim S, Kim J, Park J, Ha H, Bae JY, Kim S, Baek D (2016) General rules for functional microRNA targeting. Nature Genetics 48:1517.

\bibitem{zavo}Breda J, Rzepiela AJ, Gumienny R, van Nimwegen E, Zavolan M. Quantifying the strength of miRNA-target interactions (2015) Methods 85:90-9.

\bibitem{arve}Arvey A, Larsson E, Sander C, Leslie CS, Marks DS (2010) Target mRNA abundance dilutes microRNA and siRNA activity. Molecular systems biology 6:363.

\bibitem{levine}Levine E, Zhang Z, Kuhlman T, Hwa T (2007) Quantitative characteristics of gene regulation by small RNA. PLoS Biology 5:e229.

\bibitem{fzor}Franco-Zorrilla JM, Valli A, Todesco M, Mateos I, Puga MI, Rubio-Somoza I, Leyva A, Weigel D, Garc\'ia JA, Paz-Ares J (2007) Target mimicry provides a new mechanism for regulation of microRNA activity. Nature Genetics 39:1033.

\bibitem{salm}Salmena L, Poliseno L, Tay Y, Kats L, Pandolfi PP (2011) A ceRNA hypothesis: the Rosetta Stone of a hidden RNA language? Cell 146:353-8.

\bibitem{fati}Fatica A, Bozzoni I (2014) Long non-coding RNAs: new players in cell differentiation and development. Nature Reviews Genetics 15:7.

\bibitem{tay}Tay Y, Kats L, Salmena L, Weiss D, Tan SM, Ala U, Karreth F, Poliseno L, Provero P, Di Cunto F, Lieberman J (2011) Coding-independent regulation of the tumor suppressor PTEN by competing endogenous mRNAs. Cell 147:344-57.

\bibitem{vano}Mukherji S, Ebert MS, Zheng GX, Tsang JS, Sharp PA, van Oudenaarden A (2011) MicroRNAs can generate thresholds in target gene expression. Nature Genetics 43:854.

\bibitem{karr}Karreth FA, Tay Y, Perna D, Ala U, Tan SM, Rust AG, DeNicola G, Webster KA, Weiss D, Perez-Mancera PA, Krauthammer M (2011) In vivo identification of tumor-suppressive PTEN ceRNAs in an oncogenic BRAF-induced mouse model of melanoma. Cell 147:382-95.

\bibitem{tayy}Tay Y, Rinn J, Pandolfi PP (2014) The multilayered complexity of ceRNA crosstalk and competition. Nature 505:344.

\bibitem{yuany}Yuan Y, Liu B, Xie P, Zhang MQ, Li Y, Xie Z, Wang X (2015) Model-guided quantitative analysis of microRNA-mediated regulation on competing endogenous RNAs using a synthetic gene circuit. Proceedings of the National Academy of Sciences 112:3158-63.

\bibitem{sgro}Bosia C, Sgr\`o F, Conti L, Baldassi C, Brusa D, Cavallo F, Di Cunto F, Turco E, Pagnani A, Zecchina R (2017) RNAs competing for microRNAs mutually influence their fluctuations in a highly non-linear microRNA-dependent manner in single cells. Genome Biology 18:37.

\bibitem{stress}Leung AK, Sharp PA (2010) MicroRNA functions in stress responses. Molecular Cell 22:205-15.

\bibitem{alva}Alvarez-Garcia I, Miska EA (2005) MicroRNA functions in animal development and human disease. Development 132:4653-62.


\bibitem{anas}Anastasiadou E, Jacob LS, Slack FJ (2018) Non-coding RNA networks in cancer. Nature Reviews Cancer 18:5.

\bibitem{sanc}Sanchez-Mejias A, Tay Y (2015) Competing endogenous RNA networks: tying the essential knots for cancer biology and therapeutics. Journal of Hematology \& Oncology (2015) 8:30.

\bibitem{jens}Jens M, Rajewsky N (2015) Competition between target sites of regulators shapes post-transcriptional gene regulation. Nature Reviews Genetics 16:113.

\bibitem{denzler}Denzler R, Agarwal V, Stefano J, Bartel DP, Stoffel M (2014) Assessing the ceRNA hypothesis with quantitative measurements of miRNA and target abundance. Molecular Cell 54:766-76.

\bibitem{alau}Ala U, Karreth FA, Bosia C, Pagnani A, Taulli R, L\'eopold V, Tay Y, Provero P, Zecchina R, Pandolfi PP (2013) Integrated transcriptional and competitive endogenous RNA networks are cross-regulated in permissive molecular environments. Proceedings of the National Academy of Sciences 110:7154-9.

\bibitem{boss}Bosson AD, Zamudio JR, Sharp PA (2014) Endogenous miRNA and target concentrations determine susceptibility to potential ceRNA competition. Molecular Cell 56:347-59.

\bibitem{denz}Denzler R, McGeary SE, Agarwal V, Bartel DP, Stoffel M (2016) Impact of microRNA levels, target-site complementarity, and cooperativity on competing endogenous RNA-regulated gene expression. Molecular Cell 64:565-79.

\bibitem{wang}Wang X, Li Y, Xu X, Wang YH (2010) Toward a system-level understanding of microRNA pathway via mathematical modeling. Biosystems 100:31-8.

\bibitem{laix}Lai X, Wolkenhauer O, Vera J (2016) Understanding microRNA-mediated gene regulatory networks through mathematical modelling. Nucleic Acids Research 44:6019-35.

\bibitem{vale}Valencia-Sanchez MA, Liu J, Hannon GJ, Parker R (2006) Control of translation and mRNA degradation by miRNAs and siRNAs. Genes \& Development 20:515-24.

\bibitem{bacc}Baccarini A, Chauhan H, Gardner TJ, Jayaprakash AD, Sachidanandam R, Brown BD (2011) Kinetic analysis reveals the fate of a microRNA following target regulation in mammalian cells. Current Biology 21:369-76.

\bibitem{figl}Figliuzzi M, Marinari E, De Martino A (2013) MicroRNAs as a selective channel of communication between competing RNAs: a steady-state theory. Biophysical Journal. 104:1203-13.

\bibitem{bosi}Bosia C, Pagnani A, Zecchina R (2013) Modelling competing endogenous RNA networks. PLoS One 8:e66609.

\bibitem{meht}Noorbakhsh J, Lang AH, Mehta P (2013) Intrinsic noise of microRNA-regulated genes and the ceRNA hypothesis. PLoS One 8:e72676.

\bibitem{alon}Alon, U (2006) {\it An introduction to systems biology: design principles of biological circuits}. CRC press, Boca Raton (FL).

\bibitem{prob}Martirosyan A, Figliuzzi M, Marinari E, De Martino A (2016) Probing the limits to microRNA-mediated control of gene expression. PLoS Computational Biology 12:e1004715.

\bibitem{flon}Flondor P, Olteanu M, Stefan R (2018) Qualitative Analysis of an ODE Model of a Class of Enzymatic Reactions. Bulletin of Mathematical Biology 80:32-45.

\bibitem{kond}Sanchez A, Choubey S, Kondev J (2013) Regulation of noise in gene expression. Annual Review of Biophysics. 42:469-91.

\bibitem{vkam}Van Kampen NG (1992) {\it Stochastic processes in physics and chemistry}. Elsevier, Amsterdam.

\bibitem{swain}Swain PS (2004) Efficient attenuation of stochasticity in gene expression through post-transcriptional control. Journal of Molecular Biology 344:965-76.

\bibitem{gill}Gillespie DT (1977) Exact stochastic simulation of coupled chemical reactions. The Journal of Physical Chemistry 81:2340-61.

\bibitem{gibs}Gibson MA, Bruck J (2000) Efficient exact stochastic simulation of chemical systems with many species and many channels. The Journal of Physical Chemistry A. 104:1876-89.

\bibitem{tran}Martirosyan A, Marsili M, De Martino A (2017) Translating ceRNA susceptibilities into correlation functions. Biophysical Journal 113:206-13.

\bibitem{itza}Nitzan M, Steiman-Shimony A, Altuvia Y, Biham O, Margalit H (2014) Interactions between distant ceRNAs in regulatory networks. Biophysical journal. 106:2254-66.

\bibitem{osella11}Osella M, Bosia C, Cor\'a D, Caselle M (2011) The role of incoherent microRNA-mediated feedforward loops in noise buffering. Plos COmputational Biology 7(3): e1001101.

\bibitem{bosia12} Bosia C, Osella M, El Baroudi M, Cor\'a D, Caselle M (2012) Gene autoregulation via intronic microRNAs and its functions. BMC Systems Biology 6:131.

\bibitem{riba14}Riba A, Bosia C, El Baroudi M, Ollino L, Caselle M (2014) A Combination of Transcriptional and MicroRNA Regulation Improves the Stability of the Relative Concentrations of Target Genes. PLoS Computational Biololy 10(2):e1003490

\bibitem{osella14} Osella M, Riba A, Testori A, Cor\'a D, Caselle M (2014) Interplay of microRNA and epigenetic regulation in the human regulatory network. Frontiers in Genetics 5:345

\bibitem{grigolon16}Grigolon S, Di Patti F, De Martino A, Marinari E (2016) Noise processing by microRNA-mediated circuits: The Incoherent Feed-Forward Loop, revisited. Heliyon 2(4): e00095.

\bibitem{encode}Gerstein M, Kundaje A, Hariharan M, Landt S, Yan K, et al. (2012) Architecture of the human regulatory network derived from ENCODE data. Nature 489: 91-100.

\bibitem{bose12}Bose I, Ghosh S (2012) Origins of binary gene expression in post-transcriptional regulation by microRNAs. Eur. Phys. J. E, 35:102.

\bibitem{tsimring14}Tsimring L (2014) Noise in biology. Reports on Progress in Physics, 77:026601

\bibitem{samoilov05}Samoilov M, Plyasunov S, Arkin A (2005) Stochastic amplification and signaling in enzymatic futile cycles through noise-induced bistability with oscillations. Proceedings of the National Academy of Sciences of the U.S.A, 102(7): 2310-2315

\bibitem{delgiudice18}Del Giudice M, Bo S, Grigolon S, Bosia C (2018) On the role of microRNA-mediated bimodal gene expression. Plos Computational Biology, in press

\bibitem{lope}L\'opez-Maury L, Marguerat S, B\"ahler J (2008) Tuning gene expression to changing environments: from rapid responses to evolutionary adaptation. Nature Reviews Genetics 9:583.

\bibitem{goya}Mehta P, Goyal S, Wingreen NS (2008) A quantitative comparison of sRNA-based and protein-based gene regulation. Molecular Systems Biology 4:221.

\bibitem{cern}Martirosyan A, De Martino A, Pagnani A, Marinari E (2017) ceRNA crosstalk stabilizes protein expression and affects the correlation pattern of interacting proteins. Scientific Reports 7:43673.

\bibitem{lian}Liang H, Li WH (2007) MicroRNA regulation of human protein-protein interaction network. RNA 13:1402-8.

\bibitem{yuan}Yuan X, Liu C, Yang P, He S, Liao Q, Kang S, Zhao Y (2009) Clustered microRNAs' coordination in regulating protein-protein interaction network. BMC Systems Biology 3:65.

\bibitem{sass}Sass S, Dietmann S, Burk UC, Brabletz S, Lutter D, Kowarsch A, Mayer KF, Brabletz T, Ruepp A, Theis FJ, Wang Y (2011) MicroRNAs coordinately regulate protein complexes. BMC Systems Biology 5:136.

\bibitem{hsu}Hsu CW, Juan HF, Huang HC (2008) Characterization of microRNA-regulated protein-protein interaction network. Proteomics 8:1975-9.

\bibitem{du}Du B, Wang Z, Zhang X, Feng S, Wang G, He J, Zhang B (2014) MicroRNA-545 suppresses cell proliferation by targeting cyclin D1 and CDK4 in lung cancer cells. PloS One 9:e88022.

\bibitem{nada}Nadal A, Jares P, Pinyol M, Conde L, Romeu C, Fern\'andez PL, Campo E, Cardesa A (2007) Association of CDK4 and CCND1 mRNA overexpression in laryngeal squamous cell carcinomas occurs without CDK4 amplification. Virchows Archiv 450:161-7.

\bibitem{kwon}Kwon J, Lee TS, Lee HW, Kang MC, Yoon HJ, Kim JH, Park JH (2013) Integrin alpha 6: a novel therapeutic target in esophageal squamous cell carcinoma. International Journal of Oncology 43:1523-30.

\bibitem{tkac}Tkacik G, Callan Jr CG, Bialek W (2008) Information capacity of genetic regulatory elements. Physical Review E 78:011910.

\bibitem{dyna}Figliuzzi M, De Martino A, Marinari E (2014) RNA-based regulation: dynamics and response to perturbations of competing RNAs. Biophysical Journal 107:1011-22.

\bibitem{detw}Detwiler PB, Ramanathan S, Sengupta A, Shraiman BI (2000) Engineering aspects of enzymatic signal transduction: photoreceptors in the retina. Biophysical Journal 79:2801-17.

\bibitem{legn}Legnini I, Morlando M, Mangiavacchi A, Fatica A, Bozzoni I (2014) A feedforward regulatory loop between HuR and the long noncoding RNA linc-MD1 controls early phases of myogenesis. Molecular Cell 53:506-14.

\bibitem{fior}Fiorentino J, De Martino A (2017) Independent channels for miRNA biosynthesis ensure efficient static and dynamic control in the regulation of the early stages of myogenesis. Journal of Theoretical Biology 430:53-63.

\bibitem{maur}Mauri M, Klumpp S (2014) A model for sigma factor competition in bacterial cells. PLoS Computational Biology 10:e1003845.

\bibitem{brew}Brewster RC, Weinert FM, Garcia HG, Song D, Rydenfelt M, Phillips R (2014) The transcription factor titration effect dictates level of gene expression. Cell 156:1312-23.

\bibitem{tull}Raveh A, Margaliot M, Sontag ED, Tuller T (2016) A model for competition for ribosomes in the cell. Journal of The Royal Society Interface 13:20151062.

\bibitem{loin}Loinger A, Shemla Y, Simon I, Margalit H, Biham O (2012) Competition between small RNAs: a quantitative view. Biophysical Journal 102:1712-21.

\bibitem{koll}Koller E, Propp S, Murray H, Lima W, Bhat B, Prakash TP, Allerson CR, Swayze EE, Marcusson EG, Dean NM (2006) Competition for RISC binding predicts in vitro potency of siRNA. Nucleic Acids Research 34:4467-76.

\bibitem{nils}Pitchiaya S, Heinicke LA, Park JI, Cameron EL, Walter NG (2017) Resolving subcellular miRNA trafficking and turnover at single-molecule resolution. Cell Reports 19:630-42.

\bibitem{levlev}Levine E, McHale P, Levine H (2007) Small regulatory RNAs may sharpen spatial expression patterns. PLoS Computational Biology 3:e233.

\bibitem{levi}Teimouri H, Korkmazhan E, Stavans J, Levine E (2017) Sub-cellular mRNA localization modulates the regulation of gene expression by small RNAs in bacteria. Physical Biology 14:056001.

\end{thebibliography}
\end{document}